\documentclass[twocolumn,prl,preprintnumbers,showpacs]{revtex4}
\usepackage{amsmath}
\usepackage{amsfonts}
\usepackage{amssymb}
\usepackage{graphicx}
\usepackage{units}
\usepackage{subfigure}
\usepackage{color}
\usepackage{mathrsfs}
\usepackage{bm}
\usepackage{float}
\usepackage{pst-grad}
\usepackage{dcolumn}
\usepackage{appendix}

\newcommand{\Lim}[1]{\raisebox{0.5ex}{\scalebox{0.8}{$\displaystyle \lim_{#1}\;$}}}


\makeatletter

\begin{document}

\title{Evidence for a nematic component to the Hidden Order parameter in URu$_2$Si$_2$ from differential elastoresistance measurements}

\author{Scott C. Riggs}
\affiliation{Stanford Institute for Materials and Energy Sciences, SLAC National Accelerator Laboratory,\\ 2575 Sand Hill Road, Menlo Park, California 94025, USA} 
\affiliation{Geballe Laboratory for Advanced Materials and Department of Applied Physics, Stanford University, Stanford, California 94305, USA}
\author{M. C. Shapiro}
\affiliation{Stanford Institute for Materials and Energy Sciences, SLAC National Accelerator Laboratory,\\ 2575 Sand Hill Road, Menlo Park, California 94025, USA} 
\affiliation{Geballe Laboratory for Advanced Materials and Department of Applied Physics, Stanford University, Stanford, California 94305, USA}
\author{Akash V. Maharaj}
\affiliation{Stanford Institute for Materials and Energy Sciences, SLAC National Accelerator Laboratory,\\ 2575 Sand Hill Road, Menlo Park, California 94025, USA} 
\affiliation{Geballe Laboratory for Advanced Materials and Department of Physics, Stanford University, Stanford, California 94305, USA}
\author{S. Raghu}
\affiliation{Stanford Institute for Materials and Energy Sciences, SLAC National Accelerator Laboratory,\\ 2575 Sand Hill Road, Menlo Park, California 94025, USA} 
\affiliation{Geballe Laboratory for Advanced Materials and Department of Physics, Stanford University, Stanford, California 94305, USA}
\author{E. D. Bauer}
\affiliation{Los Alamos National Laboratory, Los Alamos, New Mexico 87545, USA} 
\author{R. E. Baumbach}
\affiliation{Los Alamos National Laboratory, Los Alamos, New Mexico 87545, USA}
\author{P. Giraldo-Gallo}
\affiliation{Stanford Institute for Materials and Energy Sciences, SLAC National Accelerator Laboratory,\\ 2575 Sand Hill Road, Menlo Park, California 94025, USA} 
\affiliation{Geballe Laboratory for Advanced Materials and Department of Physics, Stanford University, Stanford, California 94305, USA}
\author{Mark Wartenbe}
\affiliation{National High Magnetic Field Laboratory, Florida State University, Tallahassee, Florida 32310, USA} 
\author{I. R. Fisher}
\affiliation{Stanford Institute for Materials and Energy Sciences, SLAC National Accelerator Laboratory,\\ 2575 Sand Hill Road, Menlo Park, California 94025, USA} 
\affiliation{Geballe Laboratory for Advanced Materials and Department of Applied Physics, Stanford University, Stanford, California 94305, USA}

\begin{abstract}

Measurements of the differential elastoresistance of URu$_2$Si$_2$ reveal that the fluctuations associated with the 17 K Hidden Order phase transition have a nematic component.  Approaching the ``Hidden Order'' phase transition from above, the nematic susceptibility abruptly changes sign, indicating that while the Hidden Order phase has a nematic component, it breaks additional symmetries.

\end{abstract}

\maketitle

For materials that harbor a continuous phase transition, the susceptibility of the material to various ``fields'' can be used to understand the nature of the fluctuating order and hence the nature of the ordered state. In the present work we use anisotropic bi-axial strain to probe the $\it{nematic}$ susceptibility of URu$_2$Si$_2$, a heavy fermion material for which the nature of the low temperature ``Hidden Order'' state has defied comprehensive understanding for over 30 years \cite{palstra_1985, schlabitz_1986, maple_1986, JM_2011} despite considerable theoretical attention \cite{kristen_2009, varma_2006, flint_2013, das_2014, matsuda_2012, kotliar_2010, tripathi_2002, mydosh_2009}. Our measurements reveal that the Hidden Order phase has a nematic \emph{component}, confirming earlier torque measurements that reveal a broken four-fold symmetry \cite{torque_2011} and strongly constraining theoretical models describing the Hidden Order state \cite{footnote_1, taka_private}.

For a tetragonal material, the conjugate field to electronic nematic order is anisotropic biaxial in-plane strain (either $\epsilon_{aniso}=(\epsilon_{xx}-\epsilon_{yy})$ or $\epsilon_{aniso}=\gamma_{xy}$ for spontaneous order in the [100] or [110] directions respectively) \cite{footnote_2}. The nematic susceptibility for each of these orientations is then  $\chi_N \propto \frac{d \psi}{d \epsilon_{aniso}}$, where $\psi$ represents any thermodynamic quantity measuring the induced anisotropy in mutually orthogonal directions ([100] and [010], or [110] and [1$\bar{1}$0] respectively) \cite{footnote_3,JH_2012}. For a continuous electronic nematic phase transition, $\chi_N$ diverges towards the phase transition, signaling an instability towards nematic order. Such behavior was recently observed for the representative underdoped Fe-pnictide Ba(Fe$_{1-x}$Co$_x$)$_2$As$_2$, demonstrating that the structural phase transition that precedes the onset of colinear antiferromagnetic order in that material is driven by electronic nematic order \cite{JH_2012, HH_2011, HH_2013}.

There are also classes of order for which there is a single transition to an ordered state that has a nematic component, while simultaneously breaking additional symmetries. A simple example would be a unidirectional incommensurate density wave with order parameter $\Delta_{i} = \Delta(Q_{i})$, for which both translational and tetragonal symmetries are simultaneously broken at the same transition ($Q_i$ corresponds to the ordering wavevector of the density wave, and $i = x$ or $y$). In such a case, anisotropic strain is not conjugate to the order parameter (i.e., the coupling term in the free energy expansion is not bilinear). Nevertheless, the inherent nematicity associated with the unidirectional density wave motivates introduction of a nematic order parameter $\mathcal{N}$ which couples linearly to ($|\Delta_x|^2-|\Delta_y|^2$) and to which anisotropic strain \emph{is} linearly coupled. For such a situation, an analysis based on standard Ginzburg-Landau theory for coupled order parameters (see Supplemental Material \cite{SOM}) reveals that the nematic susceptibility follows a Curie-Weiss temperature dependence at high temperatures (as the density wave fluctuations orient in the anisotropic strain field) but then develops an abrupt additional divergence at the transition temperature. The additional feature arises from a symmetry-allowed contribution to the free energy of the form

\begin{equation}
\Delta F = (\epsilon_{aniso} + \lambda \mathcal{N}) (|\Delta_x|^2 - | \Delta_y|^2),
\end{equation}

\noindent where $\lambda$ is a coupling constant \cite{SOM, footnote_4}.  Within the assumption of homogeneity, such a coupling can only occur if the density wave order is described by a multicomponent order parameter \cite{footnote_5}. As we describe in detail below, we find that URu$_2$Si$_2$ appears to fall into the latter class of material, having a nematic component but also breaking additional symmetries at the phase transition.

\begin{figure}
\includegraphics[trim=0.25cm 0cm 1.75cm 1.0cm, clip=true, width=8.5cm]{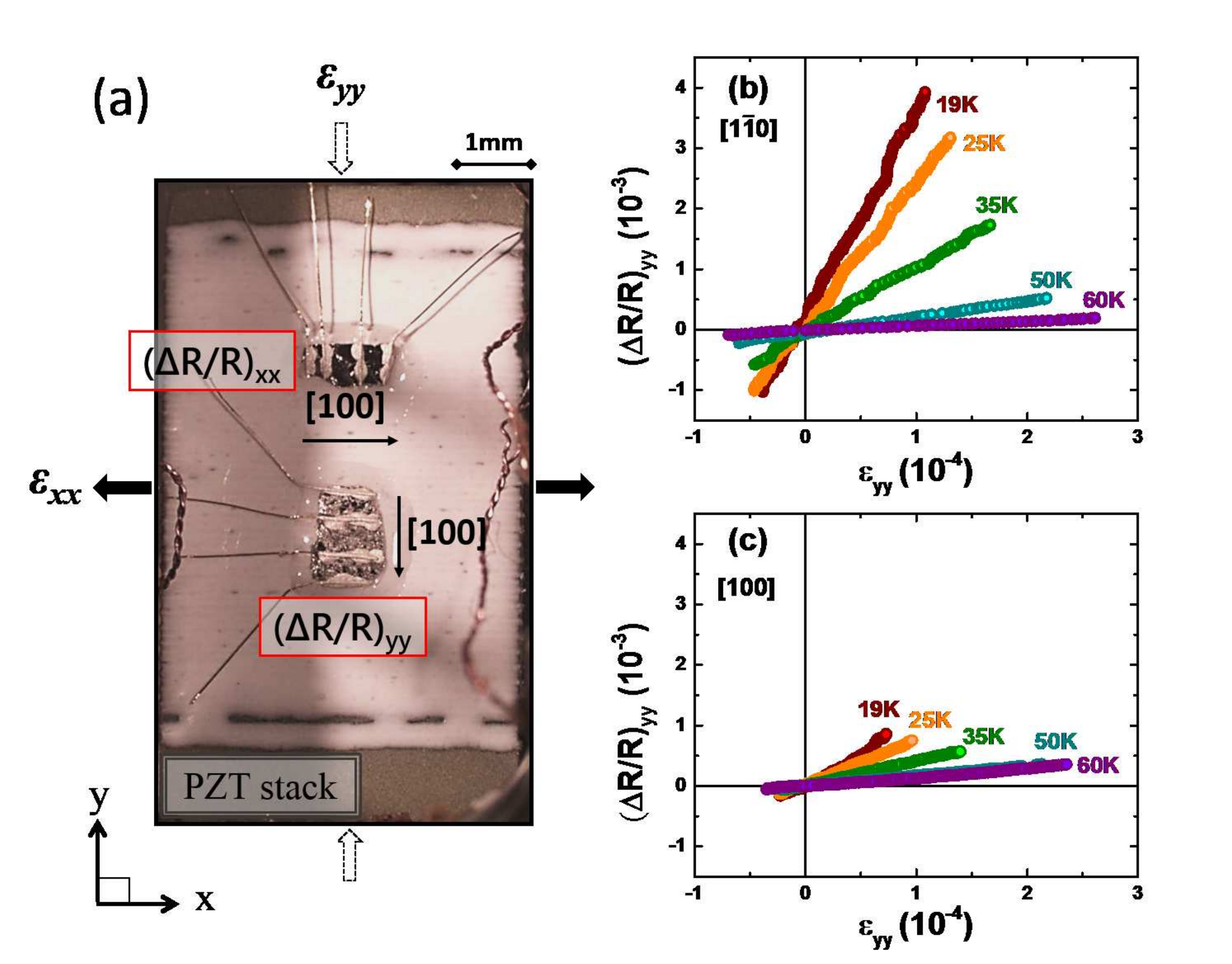} 
\caption{(Color online) Experimental measurement of elastoresitivity coefficients. (a) Photograph of [100] oriented URu$_2$Si$_2$ crystals mounted in the $(\frac{\Delta R}{R})_{yy}$ and $(\frac{\Delta R}{R})_{xx}$ directions on the surface of a PZT piezoelectric stack. Strain gauges mounted on the opposite face of the piezo stack measure $\epsilon_{yy}$ and $\epsilon_{xx}$, which are related by the effective Poisson ratio of the PZT stack, $\nu_p$. Panels (b) and (c) show representative $(\frac{\Delta R}{R})_{yy}$ elastoresistance data for five different temperatures as a function of $\epsilon_{yy}$ for [1$\bar{1}$0] and [100] orientations, respectively. Data are plotted for both warming and cooling cycles and are identical within the resolution of the experiment. The resistive response to an applied strain is considerably larger (by a factor of $\sim$4) for measurements made in the [1$\bar{1}$0] direction compared to the [100] direction.}
\label{figure_onpiezo}
\end{figure}

Our measurements are based on a novel technique that probes the differential response in the electrical resistivity to anisotropic biaxial strain in mutually perpendicular directions (Figure \ref{figure_onpiezo}). We measure the induced resistivity anisotropy, $N= \cfrac{\rho_{yy}-\rho_{xx}}{\cfrac{1}{2}(\rho_{yy}+\rho_{xx})} \sim ((\Delta R/R)_{yy}-(\Delta R/R)_{xx})$, which, by symmetry, is proportional to $\psi$ in the limit of asymptotically small values.  The induced anisotropy is characterized by specific terms in the associated elastoresistivity tensor $m_{ij}$ that relate changes in the resistivity to strains experienced by the material \cite{deltaR, SOM, HH_2013}. In the regime of infinitesimal strains (linear response), the elastoresistivity coefficients ($m_{11}-m_{12}$) and $2m_{66}$ are linearly proportional to the bare (unrenormalized) nematic susceptibility, $\chi_{N_{[100]}}$ and $\chi_{N_{[110]}}$, for strains in the [100] and [110] directions respectively \cite{HH_2013}. The proportionality constant relating the resistivity anisotropy to the nematic order parameter is governed by microscopic physics, but does not contain any singular contributions. Consequently, any divergence of the induced resistivity anisotropy is directly related to divergence of the nematic susceptibility. Since this is ultimately a derivative of a transport measurement, the technique is especially sensitive to static or fluctuating order which affects the Fermi surface.

Of particular relevance to the current experiments, recent magnetic torque measurements (measured with a small applied magnetic field in the $ab$-plane) indicated the onset of a two-fold anisotropy in the Hidden Order phase of URu$_2$Si$_2$ \cite{torque_2011}. This result is apparently supported by the observation of a weak orthorhombicity that onsets at the same temperature in high quality crystals \cite{taka_private}. These subtle effects indicate that the Hidden Order phase has a nematic component, but the degree of anisotropy that is observed depends heavily on the crystal size \cite{torque_2011} (for torque) and possibly also on the crystal quality \cite{taka_private} (for X-ray diffraction), leading to some contention as to whether or not these are intrinsic effects. By probing the temperature dependence of the nematic susceptibility in the tetragonal state, our current experiments confirm that the Hidden Order phase does indeed have a nematic component. In addition, our observation of a sharp downward deviation from mean field Curie-Weiss behavior close to $T_{HO}$ implies that the Hidden Order phase is not a pure nematic and must break additional symmetries.

\begin{figure}
\includegraphics[trim=1.5cm 0.25cm 3.cm 2.0cm, clip=true, width=8.5cm]{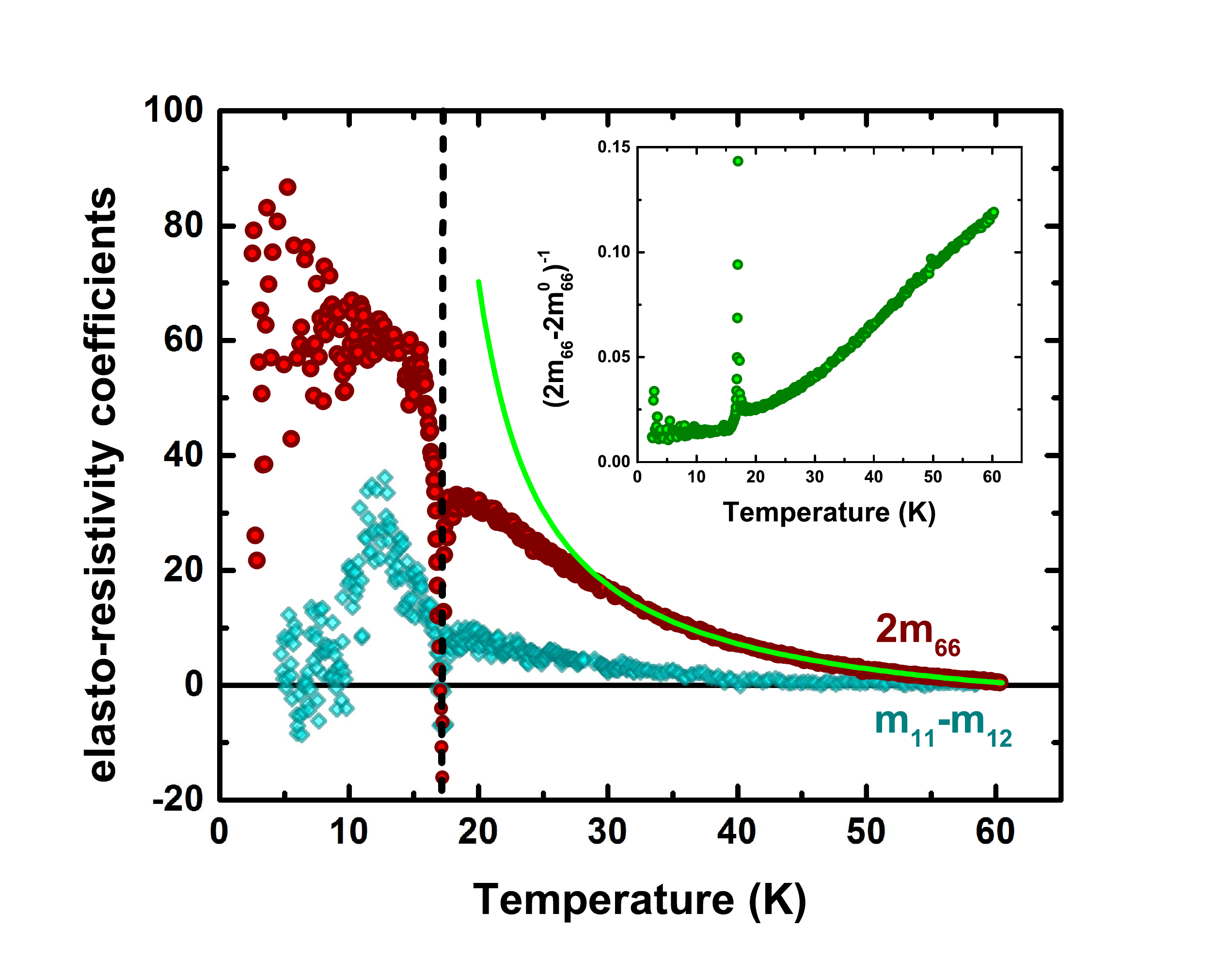} 
\caption{(Color online) Temperature dependence of the $2m_{66}$ (red circles) and the $m_{11} - m_{12}$ (cyan diamonds) elastoresistivity coefficients. The data reveal large directional anisotropy, with $2m_{66}$ reaching very large values close to and below the Hidden Order transition. The dashed vertical black line indicates the Hidden Order transition, $T_{HO} = 17.2$ K, determined from the peak in the temperature derivative of the resistivity. The green curve is a fit of $2m_{66}$ to a Curie-Weiss temperature dependence $2m_{66} = \frac{C}{T-\theta}+2m_{66}^0$ from 60 K to 30 K, as described in the main text. The fit is extrapolated below 30 K to emphasize deviations from Curie-Weiss behavior below this temperature. The values for the fitting parameters are; $C = 375$ K, $\theta = 15.2$ K, and $2m_{66}^0 = -7.9$. The inset shows $(2m_{66}-2m_{66}^0)^{-1}$ versus temperature, with $2m_{66}^0$ determined from the green fit in the main figure.}
\label{figure_elastocoeff}
\end{figure}

Anisotropic strain is achieved by gluing thin crystals of URu$_2$Si$_2$ to the surface of a PZT piezoelectric stack. The elastoresistivity coefficients $m_{11}-m_{12}$ and $2 m_{66}$ are determined from the difference of $(\Delta R/R)_{yy}$ and $(\Delta R/R)_{xx}$ for a given set of strains for [100] and [1$\bar{1}$0] oriented crystals respectively \cite{HH_2013, SOM}. To first order this is also equal to the induced resistivity anisotropy $N$, expressed below in terms of the anisotropic biaxial strain $(\epsilon_{yy}-\epsilon_{xx})$: 

\begin{equation}
N_{[100]} \approx ((\frac{\Delta R}{R})_{yy} - (\frac{\Delta R}{R})_{xx})_{[100]} = (\epsilon_{yy}-\epsilon_{xx})(m_{11} - m_{12} )
\end{equation}

\begin{equation}
N_{[110]} \approx ((\frac{\Delta R}{R})_{yy} - (\frac{\Delta R}{R})_{xx})_{[110]} = (\epsilon_{yy}-\epsilon_{xx})2 m_{66} 
\end{equation}

\noindent where $x$ and $y$ refer to the axes along which the piezo stack induces biaxial strain, indicated in Figure \ref{figure_onpiezo}(a) and following the description given in previous work \cite{HH_2011}. For each temperature, five voltage cycles were performed. The strain was measured using mutually perpendicular strain gauges glued to the back surface of the piezo stack. Representative measurements of $(\frac{\Delta R}{R})_{yy}$ for five different temperatures are shown in Figure \ref{figure_onpiezo} (b) and (c) for [1$\bar{1}$0] and [100] oriented crystals respectively, revealing a linear response. This procedure was performed for both the $(\frac{\Delta R}{R})_{xx}$ and $(\frac{\Delta R}{R})_{yy}$ orientations by removing the crystal from the piezo with acetone between each measurement. The slopes $\Delta R/R$ versus strain for each orientation were found using a linear fit and the difference was taken. The resulting elastoresistivity coefficients $2 m_{66}$ and $m_{11}-m_{12}$ are plotted in Figure \ref{figure_elastocoeff} as a function of temperature.

The data shown in Figures \ref{figure_onpiezo} and \ref{figure_elastocoeff} reveal a striking anisotropy in the elastoresistivity coefficients, with $2 m_{66}$ values considerably larger, and also more strongly temperature-dependent, than $m_{11}-m_{12}$. For comparison, the elastoresistivity coefficients of simple metals are small (of order one) and essentially isotropic \cite{HH_2013}. A sharp downward anomaly at $T_{HO}$ = 17.15 K marks the phase transition to the Hidden Order state. We return shortly to the detailed behavior in the vicinity of $T_{HO}$, first focusing on the temperature dependence of the differential elastoresistivity further above the transition. Before doing so, we note that for temperatures below approximately 15 K, the resistivity drops rapidly with decreasing temperature due to the high quality of the samples used \cite{SOM}, leading to a decrease in the signal-to-noise ratio of the measured elastoresistance (evident in the increased scatter in the $2 m_{66}$ and $m_{11}-m_{12}$ coefficients shown in Figure \ref{figure_elastocoeff}) and obscuring detailed analysis of the elastoresistance in the Hidden Order phase.

Recalling that $2 m_{66}$ is proportional to the nematic susceptibility in the [110] direction ($\chi_{N_{[110]}}$), the large values of $2 m_{66}$ (red data points in Figure \ref{figure_elastocoeff}) and the strong temperature dependence both indicate a diverging nematic susceptibility, and hence motivate fitting the data to the Curie-Weiss form ($2m_{66} = C/[(T-\theta)] + 2m_{66}^0$). Since the elastoresistance only develops significant values below approximately 60 K, the range of temperatures over which this fit can be applied is necessarily small. Nevertheless, if we attempt fits over different temperature ranges, the best fit,  defined by a reduced chi-squared statistic closest to 1, occurs for a range from approximately 30 K up to the maximum of 60 K, with a Weiss temperature $\theta =15.2 \pm 0.3$ K (see supporting material for a discussion of the fitting procedures) \cite{SOM}. Extrapolation of the best fit function to lower temperatures (shown as a green line in Figure \ref{figure_elastocoeff}) reveals significant deviation from Curie-Weiss-like behavior for temperatures below approximately 30 K. The onset of this deviation at 30 K is coincident with a suppression in $(T_{1}T)^{-1}$ seen in recent NMR experiments \cite{NMR_2013}, suggestive of a common origin, possibly associated with the progressive development of strongly fluctuating order. Perhaps coincidentally, a finite polar Kerr effect signal also develops around this temperature \cite{ES_2014}.

In contrast, $m_{11}-m_{12}$, which is proportional to $\chi_{N_{[100]}}$, is small and exhibits only a weak temperature dependence (cyan data points in Figure \ref{figure_elastocoeff}). Given the large values of $2 m_{66}$, it is likely that this weak temperature dependence derives from slight sample misalignment (i.e., a small amount of contamination from $2 m_{66}$ in the [100] orientation). For the remaining discussion we focus solely on the $2 m_{66}$ data.

\begin{figure}
\includegraphics[width=8.5cm]{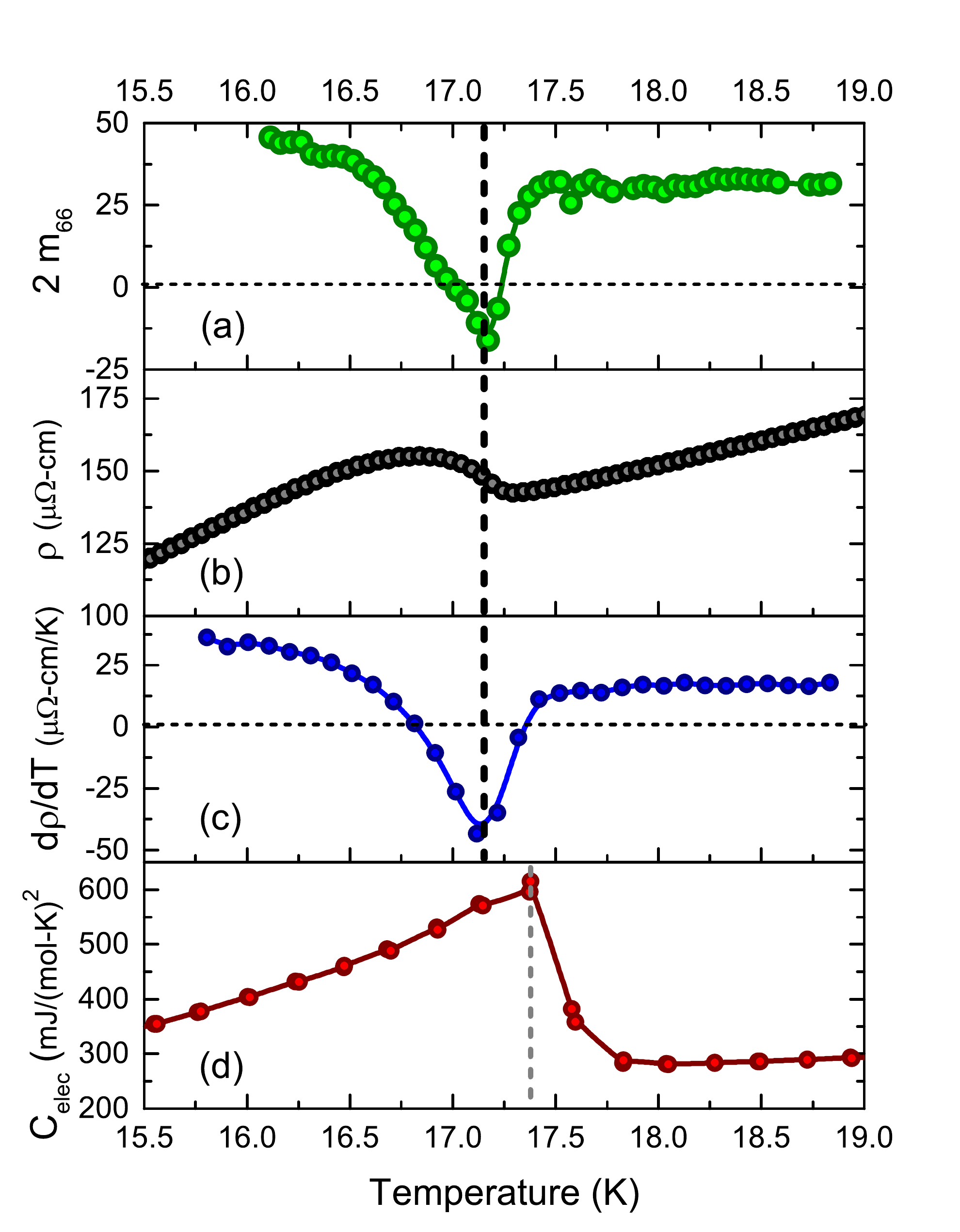} 
\caption{(Color online) Temperature dependence of (a) the $2m_{66}$ elastoresistivity coefficient, (b) the resistivity, $\rho$, (c) the temperature derivative of the resistivity, $\frac{d\rho}{dT}$, and (d) the electronic specific heat \cite{Cp_eric} in the immediate vicinity of $T_{HO}$. $T_{HO}$ obtained from the transport measurements is shown by a black vertical dashed line. Differences between T$_{HO}$ obtained from transport and electronic specific heat data (gray dashed lines) are attributed to differences in thermometry and/or RRR of the samples used for the two measurements.}
\label{figure_zoom_THO}
\end{figure}

The most striking aspect of the temperature dependence of the elastoresistivity coefficients is the sharp downward anomaly observed at $T_{HO}$. The $2m_{66}$ data are replotted in Figure \ref{figure_zoom_THO}(a) on an expanded scale for the temperature window from 15 K to 19 K, spanning $T_{HO}$. The anomaly exactly aligns with the sharp peak in the temperature derivative of the resistivity $d\rho/dT$ (Figure \ref{figure_zoom_THO}(b,c)), indicating that the effect is associated with the thermodynamic phase transition at $T_{HO}$. For comparison, heat capacity measurements were also performed for one of the crystals cleaved from the same rod that was used for the elastoresistance measurements. As has been previously established for URu$_2$Si$_2$ and as anticipated for a metallic system \cite{MF_1968}, $\frac{d\rho}{dT}$ follows the heat capacity close to $T_{HO}$, with slight differences in the peak position attributed to small differences in thermometry and residual resistivity values between the two samples.

Elastoresistance measurements for temperatures greater than $\sim$150 mK away from $T_{HO}$ reveal a linear response to strain, and the eleastoresistivity coefficients are accurately obtained from a linear fit over the entire range of strain. However, for data points within $\sim$150 mK of $T_{HO}$, the elastoresistance becomes highly nonlinear \cite{SOM}. The origin of this effect is at present unclear, but there are two natural candidates. On the one hand, the effect could arise from strain-induced changes to $T_{HO}$ \cite{JM_2011}. Within a simple mean field model of a two-component order parameter, such as would give rise to a Curie-Weiss-like nematic susceptibility, anisotropic strain results in a linear increase/decrease in $T_{HO}$ for the two distinct components of the order parameter \cite{SOM}. Estimates of $\rho(T)$ for fixed strain obtained from the elastoresistance measurements indicate a shift in $T_{HO}$ of order $\pm \sim$100 mK for strains of order $\pm 10^{-3}$, consistent with such an interpretation. Equally, as the phase transition is approached and the free energy landscape becomes progressively flatter and more strongly determined by the fluctuating order, the range of strains applied will at some point exceed the regime of linear response. In both cases, the susceptibility can still be measured by the slope of the elastoresistance at zero strain, though in practice signal-to-noise limits the range over which fits can be made. Linear fits to the elastoresistance over the full range of strain in this regime underestimate the slope at zero strain, rounding the sharp anomaly seen in Figure \ref{figure_zoom_THO} (a) and prohibiting a critical scaling analysis close to $T_{HO}$ \cite{SOM}.

The temperature dependence of the $2m_{66}$ elastoresistivity coefficient of URu$_2$Si$_2$, which is proportional to $\chi_{N_{[110]}}$, establishes that the fluctuating order associated with the Hidden Order phase has a nematic component. The progressive alignment of the fluctuating order in the anisotropic strain field yields the observed Curie-Weiss-like temperature dependence. Significantly, since $2m_{66}$ does not continue to diverge monotonically towards the eventual phase transition, but rather diverges in the $\it{opposite}$ direction as the phase transition is finally reached, it is clear that the Hidden Order phase is not a pure nematic and must break additional symmetries, as described in the Supplemental Material \cite{SOM}.  Since the $2m_{66}$ coefficient diverges but $(m_{11}-m_{12})$ does not, the orientation of the nematic fluctuations is along the [110] and [1$\bar{1}$0] directions.

The same coupling to the crystal lattice that yields the resistive response to strain must also cause an orthorhombic distortion below $T_{HO}$, consistent with reports of high resolution X-ray diffraction measurements \cite{taka_private} and also with the observation of two-fold anisotropy in the magnetic torque \cite{torque_2011}. Whether or not this anisotropy is observed in macroscopic crystals is then a matter of the resolution of the specific measurement and the relative population of the two orthogonal orthorhombic domain orientations (which will be influenced by both crystal size and quality, via effects associated with pinning of domain walls). Both aspects illustrate the distinct advantage of probing the nematic susceptibility in the tetragonal phase (i.e., for $T > T_{HO}$), for which there are no domains (one is simply probing the susceptibility of the symmetric phase) and for which the resistivity anisotropy is extremely sensitive to subtle perturbations of the Fermi surface.

Despite these advances, important questions remain. In particular, since nematic fluctuations in the $B_{2g}$ channel couple linearly to the shear modulus, one anticipates a softening of the $C_{66}$ elastic constant approaching the Hidden Order phase transition. However, earlier measurements of the elastic moduli \cite{wolf_1994, yanagisawa_2011} do not show such a softening. At least in principle, weak coupling between the nematic order parameter and the crystal lattice (consistent with the small magnitude of the orthorhombic distortion observed in X-ray diffraction experiments \cite{taka_private}) would limit any appreciable renormalization of $C_{66}$ to small values of the reduced temperature close to the phase transition \cite{footnote_6}. Nevertheless, the present results, in conjunction with other measurements that reveal a two-fold anisotropy in the Hidden Order phase \cite{torque_2011}, clearly motivate a careful reinvestigation of the elastic properties of URu$_2$Si$_2$.

In conclusion, we have shown that the $2m_{66}$ elastoresistivity coefficient of URu$_2$Si$_2$ follows a Curie-Weiss-like temperature dependence from 60 K (the highest temperature at which a finite elastoresistance can be detected by the current experimental method) down to 30 K, indicating that the flucuating order has a nematic character. As the Hidden Order phase transition is approached, the differential elastoresistance progressively deviates from the high temperature mean field behavior, and then, very close to the phase transition, develops a sharp downward divergence (Figure \ref{figure_elastocoeff}). Since the differential elastoresistance for this configuration is proportional to the nematic susceptibility $\chi_{N_{[110]}}$, this behavior directly reveals that anisotropic strain is not a conjugate field for the order parameter, and hence the phase transition must break additional symmetries beyond four-fold rotational symmetry. Taken together these data demonstrate that the Hidden Order phase has a multicomponent order parameter \cite{footnote_5}.  These results put tight theoretical constraints on any proposed theory of the Hidden Order phase transition; moreover, our measurements reveal the utility of differential elastoresistance measurements in determining the nature of subtle electronic phase transitions.

The authors thank S. A. Kivelson, Arkady Shekhter, R. Fernandes, Kristjan Haule, Gabriel Kotliar, Ross D. McDonald, B. J. Ramshaw, and Sudip Chakravarty for sitimulating discussions. The authors thank Takasada Shibauchi for sharing his data prior to submission. Work at Stanford was supported by the U.S. DOE, Office of Basic Energy Sciences, under contract DEAC02-76SF00515, and the Alfred P. Sloan Foundation (SR). Work at Los Alamos National Laboratory was performed under the auspices of the U.S. DOE, OBES, Division of Materials Sciences and Engineering.

\appendix

\section{Nematic Susceptibility: Theory}

\subsection{Electronic Nematic Transition}\label{Sec:1a}

To model a second order electronic nematic phase transition with coupling to the tetragonal lattice, one can Taylor expand the free energy to quartic order in terms of the nematic order parameter $\mathcal{N}$ and the structural order parameter $\varepsilon$:

\begin{equation}
\label{eq:1}
f = \frac{1}{2} a_0 (T - T_{\!_\mathcal{N}}) \mathcal{N}^2 + \frac{1}{4} b \mathcal{N}^4 + \frac{1}{2} c \varepsilon^2 + \frac{1}{4} d \varepsilon^4 - \lambda \mathcal{N} \varepsilon.
\end{equation}

\noindent In this expansion, the bilinear term $\lambda \mathcal{N} \varepsilon$ is the lowest order coupling allowed by symmetry, $T_{\!_\mathcal{N}}$ is the bare nematic temperature which would characterize this transition in the absence of coupling to the lattice, and $a_0, b, c, d, \lambda$ are positive coefficients.  Due to the bilinear coupling, a nonzero $\mathcal{N}$ induces a finite $\varepsilon$, which compels the $C_4 \to C_2$ point group symmetry breaking of the lattice.

In order to compute the nematic susceptibility from this free energy, we find the values of $\mathcal{N}$ which minimize $f$ by differentiating with respect to $\mathcal{N}$:

\begin{equation}
\label{eq:2}
\frac{\partial f}{\partial \mathcal{N}} = a_0 (T-T_{\mathcal{N}}) \mathcal{N} + b \mathcal{N}^3 - \lambda \varepsilon = 0.
\end{equation}

\noindent This defines a cubic equation for $\mathcal{N}$ which can be solved in terms of $\varepsilon$; however, we are interested in the linear response $\frac{\partial \mathcal{N}}{\partial \varepsilon}$, which we can find by implicit differentiation:

\begin{equation}
\label{eq:3}
\begin{split}
\frac{\partial}{\partial \varepsilon}\left(\frac{\partial f}{\partial \mathcal{N}}\right) & = [a_0 (T-T_{\mathcal{N}}) + 3b \mathcal{N}^2]\left(\frac{\partial \mathcal{N}}{\partial \varepsilon}\right) - \lambda = 0
\\ & \Rightarrow \left(\frac{\partial \mathcal{N}}{\partial \varepsilon}\right) = \frac{\lambda}{a_0 (T-T_{\mathcal{N}}) + 3b \mathcal{N}^2}.
\end{split}
\end{equation}

The nematic susceptbility is given as $\chi_{\!_\mathcal{N}} \equiv \lim_{\varepsilon \to 0} \frac{\partial \mathcal{N}}{\partial \varepsilon}$, and since $\mathcal{N} = 0$ for $T > T_{\mathcal{N}}$, above the transition we have the Curie-Weiss form:

\begin{equation}
\label{eq:4}
\chi_{\!_\mathcal{N}} = \frac{\lambda}{a_0 (T-T_{\mathcal{N}})} \quad (T > T_{\mathcal{N}}).
\end{equation}

\subsection{Nematic $\times$ (?) Symmetry}
A nematic response is also possible if the Hidden Order state breaks four-fold spatial rotation symmetry but does not couple to strain as a bilinear term in the free energy (unlike the purely electronic nematic transition in Section \ref{Sec:1a}). In this scenario, the Hidden Order state can be described by a two-component vector order parameter $\vec{\Delta}$, the nematic response can be modeled in mean field theory by introducing a separate nematic order parameter $\mathcal{N}$, and the two are coupled in the form of Equation (1) of the main text. In this section, we show that within such a theory, a sharp jump in the nematic susceptibility occurs at the Hidden Order transition and is proportional to the singular heat capacity shift of Hidden Order. This correction to the nematic susceptibility can be seen from both mean field considerations---where the jump occurs infinitesimally below the transition---and upon including Gaussian fluctuations (Sections \ref{sec:B1} and \ref{sec:B2}, respectively). 

A natural question here concerns the origin of $\mathcal{N}$. While mean field theory simply treats $\vec{\Delta}$ and $\mathcal{N}$ as coupled order parameters, and so is agnostic about their origin, it is possible that $\mathcal{N}$ is an avatar of the difference in fluctuations of the two components of $\vec{\Delta}$.  This has been argued to be the case in the iron pnictide superconductors \cite{Fernandes12}, and in Section \ref{sec:B2} we outline how the nematic order parameter arises in such a scenario. Another possibility is that there may be some intrinsic nematic ordering tendency which is preceded by the Hidden Order transition, in which case it must be included in the free energy from the start. As we show below, the free energies for these two possibilities are identical, and so the sharp discontinuity in the susceptibility occurs in either case, provided the Hidden Order transition occurs first.

\subsubsection{Mean Field Intuition}\label{sec:B1}

We consider here the mean field theory of a two-component vector order parameter $\vec{\Delta} = (\Delta_x, \Delta_y)$ coupled to a nematic order parameter $\mathcal{N}$. The minimal symmetry-allowed mean field free energy expansion takes the form
 
\begin{equation}
\label{eq:5}
\begin{split}
f & =\frac{1}{2}r_0(T-T_{HO})\left(|\Delta_x|^2+|\Delta_y|^2\right) + \frac{1}{4}u\left(|\Delta_x|^4 + |\Delta_y|^4\right) \\
&+ w|\Delta_x|^2|\Delta_y|^2 - \gamma h\left( |\Delta_x|^2-|\Delta_y|^2\right)\\ 
&+ \frac{1}{2}a_0(T-T_{\!_\mathcal{N}})\mathcal{N}^2 + \frac{1}{4}b\mathcal{N}^4 - \alpha h \mathcal{N}- \lambda \mathcal{N}\left( |\Delta_x|^2 - |\Delta_y|^2\right).
\end{split}
\end{equation}

\noindent In order that the free energy be bounded from below, we take the coefficients $a_0, b, r_0, \textrm{and } u$ to be positive, but the parameters $\alpha, \gamma, \textrm{and } \lambda$ are of indeterminate sign.  $T_{HO}$ and $T_{\!_\mathcal{N}}$ are the bare Hidden Order and nematic transition temperatures (respectively) which would characterize those phase transitions in the absence of both the external strain field $h$, and the coupling term $-\lambda \mathcal{N}\left( |\Delta_x|^2 - |\Delta_y|^2\right)$, and we assume that $T_{HO} > T_{\!_\mathcal{N}}$ so that the nematic order onsets parasitically at the Hidden Order transition. The coefficient $w$ determines whether the $Z_2$ symmetry is broken at the transition---for $w>0$ only one component orders below $T_{HO}$, while for $w<0$ both components become non-zero and there is no $Z_2$ symmetry breaking. Finally, note that we will formally take the external strain field $h$ to zero in computing the nematic susceptibility. 

We will henceforth assume that $w>0$ so that the system spontaneously breaks $Z_2$ symmetry at the transition, and we also assume that $\Delta_x$ is the component which orders (i.e. $\Delta_y = 0$ throughout). The free energy in \eqref{eq:5} then reduces to 
\begin{equation}
\label{eq:6}
\begin{split}
f & = \frac{1}{2}a_0(T-T_{\!_\mathcal{N}})\mathcal{N}^2 + \frac{1}{4}b\mathcal{N}^4 - \alpha h \mathcal{N}
\\ & + \frac{1}{2}r_0(T-T_{HO})|\Delta_x|^2 + \frac{1}{4}u|\Delta_x|^4 - \gamma h |\Delta_x|^2 - \lambda \mathcal{N} |\Delta_x|^2.
\end{split}
\end{equation}

Defining $\tilde{T}_{HO}(h,\mathcal{N}) \equiv T_{HO} + \frac{2\gamma h}{r_0} + \frac{2 \lambda \mathcal{N}}{r_0}$, we can recast \eqref{eq:6} in the more suggestive form

\begin{equation}
\label{eq:7}
\begin{split}
f & = \frac{1}{2}a_0(T-T_{\!_\mathcal{N}})\mathcal{N}^2 + \frac{1}{4}b\mathcal{N}^4 - \alpha h \mathcal{N}
\\ & + \frac{1}{2}r_0(T-\tilde{T}_{HO}(h,\mathcal{N}))|\Delta_x|^2 + \frac{1}{4}u|\Delta_x|^4.
\end{split}
\end{equation}

\noindent This form emphasizes that the effect of the $\gamma h |\Delta_x|^2$ and $\lambda \mathcal{N} |\Delta_x|^2$ terms is essentially to shift the bare Hidden Order temperature $T_{HO}$.

The phase transition occurs when $\Delta_x$ acquires a nonzero expectation value, which is found in mean field theory by computing the stationary points of $f$:

\begin{equation}
\label{eq:8}
\begin{split}
& \frac{\partial f}{\partial |\Delta_x|} = r_0(T-\tilde{T}_{HO}(h,\mathcal{N}))|\Delta_x|+u|\Delta_x|^3 = 0
\\ & \Rightarrow |\Delta_x| = \begin{cases} 0 & T > \tilde{T}_{HO}(h,\mathcal{N}) \\
\pm \sqrt{\frac{r_0(\tilde{T}_{HO}(h,\mathcal{N})-T)}{u}} & T < \tilde{T}_{HO}(h,\mathcal{N}).
\end{cases}
\end{split}
\end{equation}

\noindent Plugging these back into \eqref{eq:7}, we find that

\begin{equation}
\label{eq:9}
f = \begin{cases} f_{\textrm{nematic}} & T > \tilde{T}_{HO}(h,\mathcal{N}) \\
f_{\textrm{nematic}} - \frac{r_0^2}{4u}(\tilde{T}_{HO}(h,\mathcal{N})-T)^2 & T < \tilde{T}_{HO}(h,\mathcal{N}),
\end{cases}
\end{equation}

\noindent where $f_{\textrm{nematic}} \equiv \frac{1}{2}a_0(T-T_{\!_\mathcal{N}})\mathcal{N}^2 + \frac{1}{4}b\mathcal{N}^4 - \alpha h \mathcal{N}$ is the free energy for a second order phase transition characterized by a single nematic order parameter.  This would be the appropriate free energy to model a nematic transition in a liquid crystal, for instance, as it does not include coupling to a crystal lattice.

$\mathcal{N}$ will assume values that minimize the free energy, which we find by differentiation:

\begin{equation}
\label{eq:10}
\frac{\partial f}{\partial \mathcal{N}} = 0 = \begin{cases}
\frac{\partial f_{\textrm{nematic}}}{\partial \mathcal{N}} & T > \tilde{T}_{HO}(h,\mathcal{N}) \\
\frac{\partial f_{\textrm{nematic}}}{\partial \mathcal{N}} - \frac{r_0\lambda}{u}(\tilde{T}_{HO}-T) & T < \tilde{T}_{HO}(h,\mathcal{N})
\end{cases}
\end{equation}

\noindent with $\frac{\partial f_{\textrm{nematic}}}{\partial \mathcal{N}} = a_0(T-T_{\!_\mathcal{N}})\mathcal{N} + b\mathcal{N}^3 - \alpha h$ and $\frac{\partial \tilde{T}_{HO}}{\partial \mathcal{N}} = \frac{2\lambda}{r_0}$. \eqref{eq:10} defines a cubic equation in $\mathcal{N}$ that we could solve in terms of $h$; treating $\mathcal{N}$ as an implicit function of $h$, we can take the derivative of \eqref{eq:10} with respect to $h$ and solve for the linear response $\frac{\partial \mathcal{N}}{\partial h}$:

\begin{equation}
\label{eq:11}
\frac{\partial \mathcal{N}}{\partial h} = \begin{cases} \frac{\alpha}{a_0(T-T_{\!_\mathcal{N}})+3b\mathcal{N}^2} & T > \tilde{T}_{HO}(h,\mathcal{N}) \\
\frac{\alpha + \frac{2\gamma\lambda}{u}}{a_0(T-T_{\!_\mathcal{N}})+3b\mathcal{N}^2 - \frac{2\lambda^2}{u}} & T > \tilde{T}_{HO}(h,\mathcal{N}),
\end{cases}
\end{equation}

\noindent where we have used that $\frac{\partial \tilde{T}_{HO}}{\partial h} = \frac{2\gamma}{r_0}$.  Above and at $T = T_{HO}$, the system has not transitioned to the Hidden Order state, and hence $\Delta_x = \mathcal{N} = 0$; furthermore, the nematic susceptibility is defined in the limit of vanishing $h$.  In these limits, $\tilde{T}_{HO}(h,\mathcal{N}) = T_{HO}$ and the nematic susceptibility is given by

\begin{equation}
\label{eq:12}
\chi_{\!_\mathcal{N}} = \Lim{h \to 0} \frac{\partial \mathcal{N}}{\partial h} = \begin{cases} \frac{\alpha}{a_0(T-T_{\!_\mathcal{N}})} & T > T_{HO} \\
\frac{\alpha + \frac{2\gamma\lambda}{u}}{a_0(T-T_{\!_\mathcal{N}}) - \frac{2\lambda^2}{u}} & T < T_{HO}^{-}.
\end{cases}
\end{equation}

\noindent Note that $T_{HO}^{-}$ indicates a temperature infinitesimally below $T_{HO}$, and so there is an abrupt change in the nematic susceptibility across the transition temperature.  Above $T_{HO}$, the nematic susceptbility is exactly as for the case of an electronic nematic transition, but there are additional contributions below the ordering temperature.  This abrupt change occurs because of the change in the free energy at $T_{HO}$, which in mean field manifests as a discontinuous second derivative of the energy (or, equivalently, a discontinuity in the heat capacity).

We can relate the parameter $u$ to the magnitude of the heat capacity singularity at $T_{HO}$ by focusing on the Hidden Order contribution to the free energy on cooling through the transition and invoking the thermodynamic identity $C_v = -T\frac{\partial^2 f}{\partial T^2}$:

\begin{equation}
\label{eq:13}
\Delta C_v \equiv -T\frac{\partial^2 f}{\partial T^2}(T_{HO}^{-}) - (-T\frac{\partial^2 f}{\partial T^2}(T_{HO}^{+})) = \frac{r_0^2}{2u}T_{HO}.
\end{equation}

Using this result, we find that the nematic susceptibility in terms of the heat capacity anomaly at $T_{HO}$ is

\begin{equation}
\label{eq:14}
\chi_{\!_\mathcal{N}} = \begin{cases} \frac{\alpha}{a_0(T-T_{\!_\mathcal{N}})} & T > T_{HO} \\
\frac{\alpha}{a_0(T-T_{\!_\mathcal{N}})} \left[ 1 + \frac{4 \gamma \lambda}{r_0^2 T_{HO}} \Delta C_v + \mathcal{O}(\lambda^2) \right] & T < T_{HO}^{-}.
\end{cases}
\end{equation}

\noindent Thus, across the Hidden Order transition (where $\Delta_x$ orders), {\it the nematic susceptibility receives a correction that is proportional to the specific heat capacity jump of the Hidden Order.}  In mean field theory, this only arises just below the Hidden Order transition temperature, but fluctuations above the transition temperature have the same effect.

\subsubsection{Functional Form}\label{sec:B2}

In this section, we illustrate how a nematic order parameter can arise from fluctuations of an underlying order with $Z_2$ symmetry. Here we follow closely the procedure outlined in \cite{Fernandes12}, where we begin with a two-component vector order parameter with the free energy density

\begin{equation}
\label{eq:15}
\begin{split}
 f_1[\Delta_x, \Delta_y] &= \frac{1}{2}(\partial | \Delta_x |)^2+\frac{1}{2}(\partial | \Delta_y |)^2+\frac{r}{2}(| \Delta_x |^2+| \Delta_y |^2)
\\ &  + \frac{v}{2}(| \Delta_x |^2+| \Delta_y |^2)^2- \frac{g}{2}\left(|\Delta_x|^2-|\Delta_y|^2\right)^2\end{split}
\end{equation}
\noindent with $r \propto (T-T_c)/T_c$, $v > g$, and $\partial = \nabla$ the spatial gradient operator.  This is a simple re-writing of \eqref{eq:5} with $v=(2w+u)/4$ and $g=(2w-u)/4$.  The nematic order parameter in this situation can in fact correspond to the difference in fluctuations of the two components of $\vec{\Delta}$.  More formally, we can study such a nematic state by employing a Hubbard-Stratanovich transformation in which we decouple both fourth order terms above. We begin with the original partition function, which is a functional integral over the order parameter fields $\Delta_x, \Delta_y$:

\begin{equation}
\label{eq:16}
\begin{split}
& \mathcal{Z} \propto \int{\mathcal{D} \Delta_x \mathcal{D} \Delta_y \hspace{1mm} e^{-S_{\textrm{eff}}[\Delta_x, \Delta_y]}}
\\ & S_{\textrm{eff}}[\Delta_x, \Delta_y] = {\int{\frac{d^d\boldsymbol{k}}{(2\pi)^d} f[\Delta_x(\boldsymbol{k}),\Delta_y(\boldsymbol{k})]}}.
\end{split}
\end{equation}

\noindent In writing $S_{\textrm{eff}}$ in this way, we have assumed that $\Delta_x$ and $\Delta_y$ are frequency and momentum dependent.

To study the nematic phase, we introduce auxiliary scalar fields $\mathcal{N}$ for $| \Delta_x |^2 - | \Delta_y |^2$ and $\zeta$ for $| \Delta_x |^2 + | \Delta_y |^2$ and perform a Hubbard-Stratonovich transformation to decouple the quartic order (in $\Delta_i$) terms in the free energy.  The Hubbard-Stratonovich transformation is a generalization of the identity

\begin{equation}
\label{eq:17}
e^{-\frac{a}{2}x^2} = \frac{1}{\sqrt{2 \pi a}} \int_{-\infty}^{\infty}{dy \hspace{1mm} e^{-\frac{y^2}{2a}-ixy}},
\end{equation}

\noindent which for separate fields $\mathcal{N}$ and $\zeta$ yields

\begin{equation}
\label{eq:18}
\begin{split}
& \mathcal{Z} \propto \int{\mathcal{D} \Delta_x \mathcal{D} \Delta_y \mathcal{D} \mathcal{N} \mathcal{D} \zeta \hspace{1mm} e^{-S_{\textrm{eff}}[\Delta_x, \Delta_y, \mathcal{N}, \zeta]}}
\\ & S_{\textrm{eff}}[\Delta_x, \Delta_y, \mathcal{N}, \zeta] ={\int{\frac{d^d\boldsymbol{k}}{(2\pi)^d} f_2[\Delta_x(\boldsymbol{k}),\Delta_y(\boldsymbol{k}),\mathcal{N}(\boldsymbol{k}),\zeta(\boldsymbol{k})]}}.
\end{split}
\end{equation}

\noindent This represents a major simplification because we have reduced the theory to one which includes only Gaussian integrals, but this comes at the cost of two additional integrals over the extra auxiliary order parameter fields.

We would like to compute the nematic susceptibility above the temperature at which both $\vec{\Delta}$ and $\mathcal{N}$ order (i.e., $\langle \Delta_x \rangle = \langle \Delta_y \rangle = \langle \mathcal{N} \rangle = 0$), and so, in addition to transforming the free energy in \eqref{eq:5}, we also add in the effect of a symmetry breaking strain field $h$. Proceeding to Gaussian level in the order parameter fields and the fluctuations, this corresponds to the theory

\begin{equation}
\label{eq:19}
\begin{split}
&f_2[\Delta_x(\mathbf{r}), \Delta_y(\mathbf{r}), \mathcal{N}, \zeta] = \frac{1}{2}(\partial |\Delta_x|)^2 + \frac{1}{2}(\partial |\Delta_y|)^2
\\&+ \frac{r}{2}(|\Delta_x|^2 + |\Delta_y|^2)  -  \frac{h}{2}(|\Delta_x|^2 - |\Delta_y|^2)
\\& +\frac{1}{2}a\mathcal{N}^2 + \lambda (|\Delta_x|^2 - |\Delta_y|^2) \mathcal{N}- \alpha h \mathcal{N}
\\ & +\frac{1}{2}b \zeta^2 + d(|\Delta_x|^2 + |\Delta_y|^2)\zeta ,
\end{split}
\end{equation}

\noindent where $r \propto (T-T_c)/T_c$ and $a = \frac{1}{g} \propto (T-T_{\!_\mathcal{N}})/T_{\!_\mathcal{N}} $ (Note: we assume that $T_c > T_{\!_\mathcal{N}}$ so that nematic order develops parasitically when $\vec{\Delta}$ orders).  The coefficient $\alpha$ reflects the  strength and sign of the coupling of $\mathcal{N}$ to the external field, relative to that of $\vec{\Delta}$. 

The field $\zeta$ (with quadratic coefficient $b=-1/v$) essentially renormalizes the coefficient $r$ in \eqref{eq:19}, and so we neglect its fluctuations and analyze the theory by integrating out $\Delta_x$ and $\Delta_y$:

\begin{equation}
\label{eq:20}
\begin{split}
\mathcal{Z} & \propto \int{\mathcal{D} \mathcal{N} \hspace{1mm} \left\langle e^{-\int_k{\lambda (|\Delta_x|^2 - |\Delta_y|^2) \mathcal{N}}} \right\rangle_{\Delta_x,\Delta_y} e^{-\int_k{\frac{a}{2}\mathcal{N}^2-\alpha h \mathcal{N}}}}
\\ & \approx \int{\mathcal{D} \mathcal{N} \hspace{1mm} e^{-\int_k{\lambda (\langle|\Delta_x|^2\rangle - \langle|\Delta_y|^2\rangle) \mathcal{N}}} e^{-\int_k{\frac{a}{2}\mathcal{N}^2-\alpha h \mathcal{N}}}},
\end{split}
\end{equation}

\noindent with $\langle |\Delta_x|^2 \rangle$ given by

\begin{equation}
\label{eq:21}
\begin{split}
& \int\mathcal{D} \Delta_x \,\,(\Delta_x^2) \exp \left[ -\frac{1}{2} \int_{\boldsymbol{k}}\left(\boldsymbol{k}^2+r+h\right)\Delta_x(\boldsymbol{k})\Delta_x(\boldsymbol{-k}) \right]
\\ & = \int{\frac{d^{d}\boldsymbol{k}}{(2\pi)^{d}}\frac{1}{[\boldsymbol{k}^2+(r/2)+h]}} \equiv G(r+h)
\end{split}
\end{equation}

\noindent and $\langle |\Delta_y|^2 \rangle$ given by

\begin{equation}
\label{eq:22}
\begin{split}
& \int\mathcal{D} \Delta_y\,\, (\Delta_y^2) \exp \left[ -\frac{1}{2} \int_{\boldsymbol{k}}\left(\boldsymbol{k}^2+r-h\right)\Delta_y(\boldsymbol{k})\Delta_y(\boldsymbol{-k}) \right]
\\ & = \int{\frac{d^{d}\boldsymbol{k}}{(2\pi)^{d}}\frac{1}{[\boldsymbol{k}^2+(r/2)-h]}} \equiv G(r-h).
\end{split}
\end{equation}

Thus, taking the saddle point of the partition function in \eqref{eq:20}, we have for small values of the applied field

\begin{equation}
\label{eq:23}
\begin{split}
& a\mathcal{N}-\alpha h+\lambda[G(r+h)-G(r-h)] = 0
\\ & \Longrightarrow a\mathcal{N}-\alpha h+\lambda \left( \frac{\partial G(r)}{\partial r} \right)h = 0.
\end{split}
\end{equation}

\noindent This in turn implies that the nematic susceptibility above $T_c$ is given by

\begin{equation}
\label{eq:24}
\begin{split}
\chi_{\!_\mathcal{N}} & = \lim_{h \to 0} \frac{\partial \mathcal{N}}{\partial h} = \frac{\alpha}{a} \left[1 - \lambda \left( \frac{\partial G(r)}{\partial r} \right)  \right]
\\ & = \frac{\alpha}{a_0(T-T_{\!_\mathcal{N}})} \left[1 - \lambda \left( \frac{\partial G(r)}{\partial r} \right)  \right] \quad (T > T_c).
\end{split}
\end{equation}

\noindent  What \eqref{eq:24} shows is that in the disordered phase of $\vec{\Delta}$, the nematic susceptibility above the phase transition is the sum of a Curie-Weiss term (the same as for the electronic nematic case) and an additional contribution which, as will be shown below, has a specific heat-like dependence on temperature.

\subsubsection{Specific Heat-like Singularity}

To see why $\frac{\partial G(r)}{\partial r}$ behaves like the specific heat singularity due to the phase transition, we will express the partition function in terms of the mean field order parameters (including spatial fluctuations), derive the free energy $f$, and then compute the heat capacity singularity due to the phase transition as $c_{\textrm{sing}} = \frac{\partial E}{\partial T} = \frac{\partial}{\partial T} \left[- \frac{\partial \ln \mathcal{Z}}{\partial \beta} \right] = \frac{\partial}{\partial T} \left[ \frac{\partial f}{\partial \beta} \right]$.  Although we focus only on the heat capacity contribution due to the phase transition $c_{\textrm{sing}}$, there are many other degrees of freedom that contribute to the total heat capacity; however, these other degrees of freedom do not change abruptly at $T_c$.

Following this procedure, we consider the full partition function (with fluctuations) up to quartic order in the $\vec{\Delta}$ fields:

\begin{equation}
\label{eq:25}
\begin{split}
& \mathcal{Z} \propto \int{\mathcal{D} \Delta_x \mathcal{D} \Delta_y \hspace{1mm} e^{-S_{\textrm{eff}}[\Delta_x, \Delta_y]}}
\\ & S_{\textrm{eff}}[\Delta_x, \Delta_y] = \int{\frac{d^d\boldsymbol{k}}{(2\pi)^d} f[\Delta_x(\boldsymbol{k}),\Delta_y(\boldsymbol{k})]}
\\ & f[\Delta_x, \Delta_y] = \frac{1}{2}(\partial | \Delta_x |)^2+\frac{1}{2}(\partial | \Delta_y |)^2+\frac{r}{2}(| \Delta_x |^2+| \Delta_y |^2)
\\ & \hspace{10mm} + \frac{u}{4}(| \Delta_x |^4+| \Delta_y |^4)+ w|\Delta_x|^2|\Delta_y|^2
\end{split}
\end{equation}

We begin by expanding $\vec{\Delta}$ about its mean field value, i.e., by letting

\begin{equation}
\label{eq:26}
\begin{split}
& |\Delta_x| = \langle |\Delta_x| \rangle + \psi_x(\boldsymbol{r})
\\ & |\Delta_y| = \langle |\Delta_y| \rangle + \psi_y(\boldsymbol{r}),
\end{split}
\end{equation}

\noindent where the mean field values of the components of the order parameter are given by

\begin{equation}
\label{eq:27}
\langle |\Delta_x| \rangle = \begin{cases}
0 & \text{if $T > T_c$} \\
\pm \sqrt{-|r|/u} & \text{if $T < T_c$},
\end{cases}
\end{equation}

\noindent and $\langle |\Delta_y| \rangle = 0$ for all temperatures (since the state is unidirectional).  Expanding the terms in \eqref{eq:25} to quadratic order in the fluctuations $\psi_x$ and $\psi_y$,

\begin{equation}
\label{eq:28}
\begin{split}
& \frac{1}{2}(\partial |\Delta_x|)^2 = \frac{1}{2}(\partial[\langle |\Delta_x| \rangle + \psi_x])^2 = \frac{1}{2}(\nabla \psi_x)^2
\\ & \frac{1}{2}(\partial |\Delta_y|)^2 = \frac{1}{2}(\nabla \psi_y)^2
\\ & \frac{r}{2}(| \Delta_x |^2+| \Delta_y |^2) = \frac{r}{2}(\langle | \Delta_x | \rangle ^2 + \psi_x^2 + 2\langle | \Delta_x | \rangle \psi_x + \psi_y^2)
\\ & \frac{u}{4}(|\Delta_x|^4+|\Delta_y|^4) \approx \frac{u}{4} [ \langle |\Delta_x| \rangle ^4 + 4\langle |\Delta_x| \rangle ^3 \psi_x
+ 6\langle |\Delta_x| \rangle ^2 \psi_x^2  ]
\\ & w|\Delta_x|^2|\Delta_y|^2 \approx w \langle |\Delta_x| \rangle ^2 \psi_y^2.
\end{split}
\end{equation}

Observing that the linear terms in $\psi_x$ cancel, the new theory including fluctuations is given by

\begin{equation}
\label{eq:29}
\begin{split}
& \mathcal{Z} \propto \int{\mathcal{D} \psi_x \mathcal{D} \psi_y \hspace{1mm} e^{-S_{\textrm{eff}}[\psi_x, \psi_y]}}
\\ & \frac{1}{\mathcal{V}} S_{\textrm{eff}} = \frac{r}{2} \langle |\Delta_x| \rangle ^2 + \frac{u}{4} \langle |\Delta_x| \rangle ^4
\\ & \hspace{10mm} + \frac{1}{2\mathcal{V}} \int{\frac{d^d\boldsymbol{k}}{(2\pi)^d}(\boldsymbol{k}^2+\xi_x^{-2})|\psi_x(\boldsymbol{k})|^2}
\\ & \hspace{10mm} + \frac{1}{2\mathcal{V}} \int{\frac{d^d\boldsymbol{k}}{(2\pi)^d}(\boldsymbol{k}^2+\xi_y^{-2})|\psi_y(\boldsymbol{k})|^2}
\\ & \equiv \frac{r}{2} \langle |\Delta_x| \rangle ^2 + \frac{u}{4} \langle |\Delta_x| \rangle ^4 
\\ & \hspace{10mm} + \frac{1}{2\mathcal{V}} \int{\frac{d^d\boldsymbol{k}}{(2\pi)^d} [G_x^{-1}(\boldsymbol{k})|\psi_x(\boldsymbol{k})|^2+G_y^{-1}(\boldsymbol{k})|\psi_y(\boldsymbol{k})|^2}],
\end{split}
\end{equation}

\noindent where we have introduced the length scales

\begin{equation}
\label{eq:30}
\begin{split}
\xi_x^{-2} & \equiv \frac{1}{2}(r+3u\langle |\Delta_x| \rangle^2) = \begin{cases}
r/2 & T > T_c \\
|r| & T < T_c
\end{cases}
\\ \xi_y^{-2} & \equiv \frac{1}{2}(r+2w\langle |\Delta_x| \rangle^2) = \begin{cases}
r/2 & T > T_c \\
|r|(2w-u)/2u & T < T_c.
\end{cases}
\end{split}
\end{equation}

Using the formula $\ln \det G^{-1} =  \textrm{Tr} \ln G^{-1}$, we can integrate out the fluctuation fields $\psi_x, \psi_y$ in \eqref{eq:29} to obtain the free energy density:

\begin{equation}
\label{eq:31}
\begin{split}
f = -\frac{\ln \mathcal{Z}}{\mathcal{V}} = & \frac{r}{2}\langle |\Delta_x| \rangle^2 + \frac{u}{4}\langle |\Delta_x| \rangle^4
\\ & + \frac{1}{2} \int{\frac{d^d\boldsymbol{k}}{(2\pi)^d} (\ln [\boldsymbol{k}^2+\xi_x^{-2}] +\ln [\boldsymbol{k}^2+\xi_y^{-2}])}.
\end{split}
\end{equation}

\noindent For $T > T_c$, the two terms in the integral are the same.

We can now compute the heat capacity singularity $c_{\textrm{sing}} = \frac{\partial}{\partial T} \left[ \frac{\partial f}{\partial \beta} \right]$.  Since $r \propto (T-T_c)/T_c$, it follows that $\frac{\partial}{\partial T} = \frac{\partial r}{\partial T}\frac{\partial}{\partial r} = \frac{1}{T_c}\frac{\partial}{\partial r}$ and $\frac{\partial}{\partial \beta} = -T^2\frac{\partial}{\partial T} = \frac{-T^{2}_c(1+r)^2}{T_c}\frac{\partial}{\partial r} \approx -T_c\frac{\partial}{\partial r}$ (valid for $T \approx T_c$, or $|r| \ll 1$), and hence the heat capacity singularity is given by

\begin{equation}
\label{eq:32}
c_{\textrm{sing}} = -\frac{\partial^2 f}{\partial r^2}.
\end{equation}

\noindent Applying this result to \eqref{eq:31} and substituting the $r$ dependence of $\langle |\Delta_x| \rangle$, we find that for $T > T_c$

\begin{equation}
\label{eq:33}
c_{\textrm{sing}} = \frac{\partial}{\partial r} \left[ \int{\frac{d^d\boldsymbol{k}}{(2\pi)^d}\frac{1}{[\boldsymbol{k}^2+(r/2)]}} \right] \equiv \frac{\partial G(r)}{\partial r} \quad (T > T_c),
\end{equation}

\noindent where $G(r)$ is the correlation function that we wrote down in the previous section. Hence, $\partial G/\partial r$ is equal to the fluctuational contribution to the heat capacity and proportional to the nematic susceptibility correction in the disordered phase.

\section{Elastoresistivity Measurements}

For a tetragonal material, the elastoresistivity tensor (which relates the normalized resistivity change to the strain in different directions) is defined by

$\scriptscriptstyle{\begin{pmatrix} \left( \nicefrac{\Delta \rho}{\rho} \right)_{xx} \\ \left( \nicefrac{\Delta \rho}{\rho} \right)_{yy} \\ \left( \nicefrac{\Delta \rho}{\rho} \right)_{zz} \\ \left( \nicefrac{\Delta \rho}{\rho} \right)_{yz} \\ \left( \nicefrac{\Delta \rho}{\rho} \right)_{zx} \\ \left( \nicefrac{\Delta \rho}{\rho} \right)_{xy} \end{pmatrix} = \begin{pmatrix} m_{11} & m_{12} & m_{13} & 0 & 0 & 0 \\ m_{12} & m_{11} & m_{13} & 0 & 0 & 0 \\ m_{13} & m_{13} & m_{33} & 0 & 0 & 0 \\ 0 & 0 & 0 & m_{44} & 0 & 0 \\ 0 & 0 & 0 & 0 & m_{44} & 0 \\ 0 & 0 & 0 & 0 & 0 & m_{66} \\ \end{pmatrix} \begin{pmatrix} \epsilon_{xx} \\ \epsilon_{yy} \\ \epsilon_{zz} \\ \epsilon_{yz} \\ \epsilon_{zx} \\ \epsilon_{xy} \end{pmatrix}}$.

\noindent In writing this equation, we have employed the Voigt notation, which allows us to express the rank 4 elastoresistivity tensor as a matrix and the rank 2 strain and resistivity change tensors as 6-component vectors.  Crucially for our experiment, we define our Cartesian axes relative to the piezo stack itself, with the x-(y-)direction parallel to the short (long) axis of the piezo.

The elastoresistivity coefficients are determined by measuring the resistance change in response to an applied strain. A sample is mounted to a piezo stack with a thin layer of Devcon 5-min epoxy. Applying a positive voltage to the piezo stack causes it to expand in the longitudinal direction (long axis of the piezo) and contract in the transverse direction (short axis of the piezo). The ratio of the longitudinal expansion, which applies positive $\epsilon_{yy}$ strain on the sample, to the transverse contraction which applies $-\epsilon_{xx}$ strain to the sample, is the Poisson ratio of the piezo stack, $\nu_p = -\frac{\epsilon_{xx}}{\epsilon_{yy}}$. Note that this convention for the piezo stack Poisson ratio is the inverse of that defined in our previous work \cite{HH_2013}. For thin samples mounted on the piezo stack the in-plane sample strain is equal to the piezo strain. The out-of-plane sample strain is free to expand or contract and is controlled by the Poisson ratio of the material $\nu_s$. The measured change in resistance to applied strain from the piezo stack is therefore an admixture of several elastoresistivity coefficients. To disentangle particular elastoresistivity coefficients, measurements are made in the longitudinal $(\frac{\Delta R}{R})_{yy}$ and transverse $(\frac{\Delta R}{R})_{xx}$ configurations along high symmetry directions. Specifically, measurements are made for crystals oriented such that the current flows along the [100] direction and the [110] directions \cite{HH_2013}. Differential elastoresistance measurements for these two configurations yields:

\begin{equation}
\label{eq:34}
\begin{split}
N_{[100]} & = ((\Delta R/R)_{yy} - (\Delta R/R)_{xx})_{[100]}
\\ & = \epsilon_{yy}(1+ \nu_p)(m_{11} - m_{12} )
\end{split}
\end{equation}

\begin{equation}
\label{eq:35}
\begin{split}
N_{[110]} & = ((\Delta R/R)_{yy} - (\Delta R/R)_{xx})_{[110]}
\\ & = \epsilon_{yy}(1+ \nu_p)2 m_{66}
\end{split}
\end{equation}

The above two equations are used to extract $m_{11} - m_{12}$ and $2m_{66}$, which correspond to the B$_{1g}$ and B$_{2g}$ symmetry channels respectively. As explained in the main text, these coefficients are proportional to the components of the nematic susceptibility tensor $\chi_{\!_{N_{[100]}}}$ and $\chi_{\!_{N_{[110]}}}$, relevant to nematic order in the [100] and [110] directions, respectively.

\section{URu$_2$Si$_2$ Specifics}

\subsection{Crystal Growth}

\begin{figure}
\includegraphics[width=8.5cm]{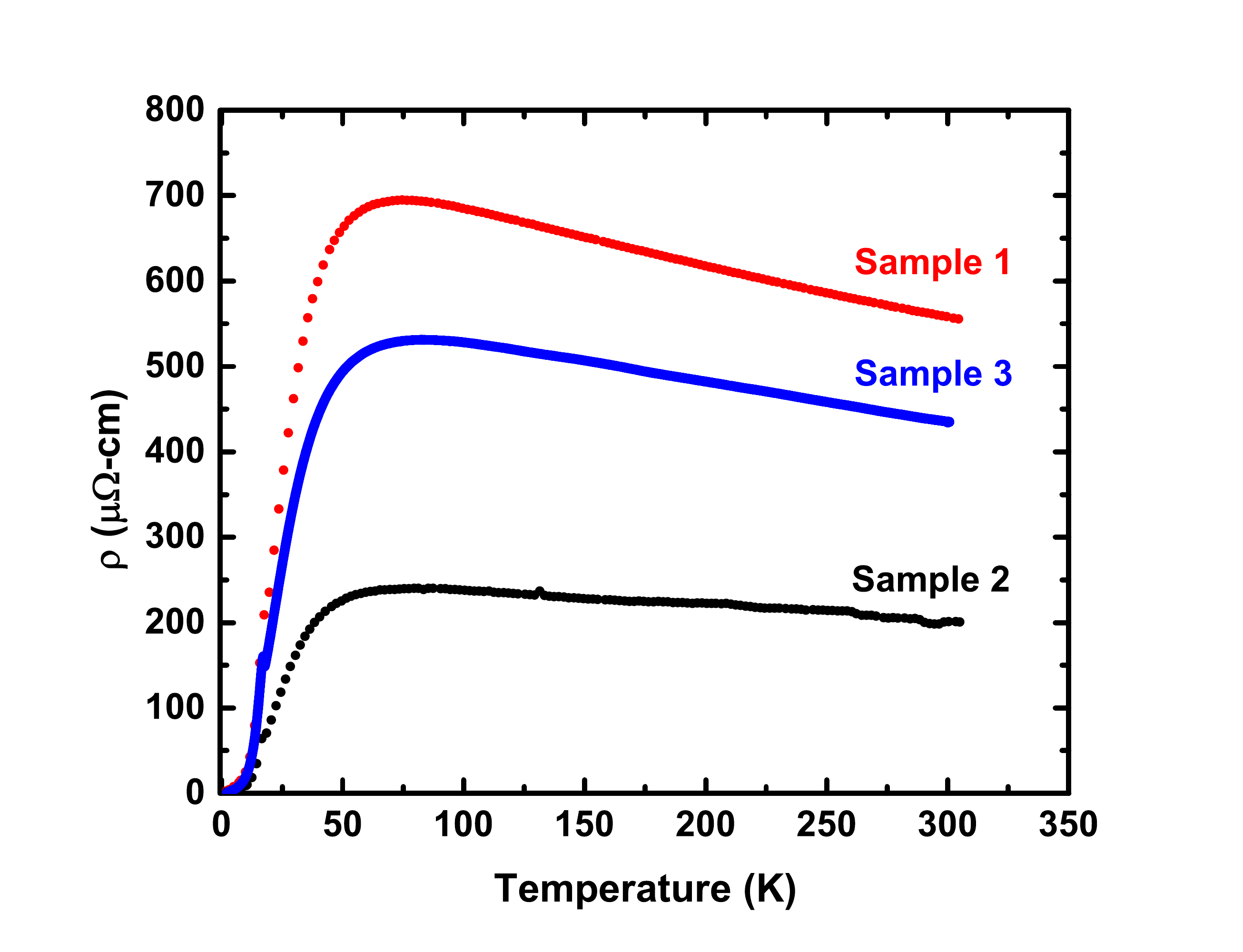} 
\caption{(Color online) Resistivity as a function of temperature for samples detailed in table \ref{table1}. Sample 1 [100] and Sample 3 [1$\bar{1}$0] were used for the elastoresistivity measurements in the main text. Sample 2 [100] was also measured and showed the same temperature dependence in $m_{11}-m_{12}$ as Sample 1.}
\label{figure_rhovT}
\end{figure}

Single crystals of URu$_2$Si$_2$ were grown by the Czochralski technique and then electro-refined to improve the purity.  The crystals were oriented by Laue diffraction in back-scattering geometry and cut with a wire saw.  In some cases, the crystals were cleaved with a razor blade and bar shaped samples were extracted. After elastoresitivity measurements the in-plane orientations of the URu$_2$Si$_2$ samples were confirmed using a commercial four-circle single crystal diffractometer in an Eulerian cradle geometry, with CuK$_{\alpha 1}$ radiation, of wavelength $\lambda = 1.54 \AA$. Fine alignment of the diffraction geometry was done using the (004) out-of-plane orientation peak, which is the one with the strongest intensity for this family of planes. Subsequently, the [112] peaks were identified by moving to the corresponding $\omega$ and $\xi$ (azimuthal) positions, and performing $\phi$ scans (rotation about the out-of-plane axis). $\phi$=0 corresponds to top and bottom sample edges along the horizontal axis. For crystals with edges along the [100] directions, the [112] diffraction peaks were found at $\phi$ values of 45$^{\circ}$, 135$^{\circ}$, 225$^{\circ}$ and 315$^{\circ}$. For crystals with edges along the [110] directions, the [112] diffraction peaks were found at $\phi$ values of 0$^{\circ}$, 90$^{\circ}$, 180$^{\circ}$ and 270$^{\circ}$. Table \ref{table1} provides details for the three samples measured, where $RRR = \frac{R(300K)}{R(4.2K)}$. The resistivity as a function of temperature for the three samples is plotted in Figure \ref{figure_rhovT}.

\begin{table}[h]
\small
\caption{URu$_2$Si$_2$ Sample Information (in mm)}
\centering
\begin{tabular}{ c c c c c c }
\hline\hline
  Sample & direction & length & width & thickness & RRR \\
	Sample 1 & [100] & 1.05 & 0.073 & 0.070 & 87\\
	Sample 2 & [100] & 1.38 & 0.656 & 0.062 & 81\\
	Sample 3 & [1$\bar{1}$0] & 1.26 & 0.288 & 0.095 & 170\\
  
\hline
\end{tabular}
\label{table1}
\end{table}

\subsection{Proximity to Strain-induced Transition}

It has previously been established that hydrostatic pressure in excess of approximately 5 kBar stabilizes the competing local moment antiferromagnetic (LMA) groundstate of URu$_2$Si$_2$ (for a relatively recent review of key experimental results see \cite{JM_2011} and references therein). In addition, recent theoretical treatment employing a Ginzburg-Landau model based on the lowest-lying crystal field states predicts a similar stabilization of the LMA state due to uniaxial stress \cite{kotliar_2010}. In this theory, the HO-to-LMA transition is predicted to occur for uniaxial stress $\sigma \sim 0.6$ GPa directed along either the [100] or [110] directions. Although this effect has yet to be established experimentally, it is important to compare the strains experienced by the crystals in our experiment, with those that would be anticipated for such large stresses. Since the elastic moduli of URu$_2$Si$_2$ are known \cite{wolf_1994}, we can readily calculate the strain that would result from uniaxial stress of this magnitude. In the following analysis we show that our experiment employs strains that are considerably smaller than those corresponding to the critical stress boundaries.  

For tetragonal symmetry, strain and stress are related by the tensor equation

$ \scriptscriptstyle{\begin{pmatrix} \sigma_{xx} \\ \sigma_{yy} \\ \sigma_{zz} \\ \sigma_{yz} \\ \sigma_{zx} \\ \sigma_{xy} \end{pmatrix}= \begin{pmatrix} C_{11} & C_{12} & C_{13} & 0 & 0 & 0 \\ C_{12} & C_{11} & C_{13} & 0 & 0 & 0 \\ C_{13} & C_{13} & C_{33} & 0 & 0 & 0 \\ 0 & 0 & 0 & C_{44} & 0 & 0 \\ 0 & 0 & 0 & 0 & C_{44} & 0 \\ 0 & 0 & 0 & 0 & 0 & C_{66} \\ \end{pmatrix} \begin{pmatrix} \epsilon_{xx} \\ \epsilon_{yy} \\ \epsilon_{zz} \\ \epsilon_{yz} \\ \epsilon_{zx} \\ \epsilon_{xy} \end{pmatrix}}$.

\noindent
In writing this tensor equation we have employed the Voigt notation convention, whereby $xx \rightarrow 1$, $yy \rightarrow 2$, $zz \rightarrow 3$, $yz \rightarrow 4$, $zx \rightarrow 5$, and $xy \rightarrow 6$. Answering our main question requires inverting the elastic moduli tensor, $C_{ij}$, which for tetragonal symmetry gives

$ \scriptscriptstyle{\begin{pmatrix} \epsilon_{xx} \\ \epsilon_{yy} \\ \epsilon_{zz} \\ \epsilon_{yz} \\ \epsilon_{zx} \\ \epsilon_{xy} \end{pmatrix}=}$ 

$\scriptscriptstyle{\begin{pmatrix} \frac{C_{11}C_{33}-C_{13}^2}{\alpha(C_{11}-C_{12})} & \frac{C_{13}^2-C_{12}C_{33}}{\alpha(C_{11}-C_{12})} & -\frac{C_{13}}{\alpha} & 0 & 0 & 0 \\ \frac{C_{13}^2-C_{12}C_{33}}{\alpha(C_{11}-C_{12})} & \frac{C_{11}C_{33}-C_{13}^2}{\alpha(C_{11}-C_{12})} & -\frac{C_{13}}{\alpha} & 0 & 0 & 0 \\ -\frac{C_{13}}{\alpha} & -\frac{C_{13}}{\alpha} & \frac{C_{11}+C_{12}}{\alpha} & 0 & 0 & 0 \\ 0 & 0 & 0 & \frac{1}{C_{44}} & 0 & 0 \\ 0 & 0 & 0 & 0 & \frac{1}{C_{44}} & 0 \\ 0 & 0 & 0 & 0 & 0 & \frac{1}{C_{66}} \\ \end{pmatrix} \begin{pmatrix} \sigma_{xx} \\ \sigma_{yy} \\ \sigma_{zz} \\ \sigma_{yz} \\ \sigma_{zx} \\ \sigma_{xy} \end{pmatrix}}$.

\noindent
where $\alpha \equiv C_{33}(C_{11}+C_{12})-2C_{13}^2$.  Explicit inversion of this matrix was done in Mathematica, but the exact form is also given in \cite{Ballato95}.

In this form, we can now consider the cases of a uniaxial stress of magnitude $\sigma$ applied in the [100] and [110] directions.  Starting with the simpler case, a uniaxial stress of magnitude $\sigma$ applied in the [100] direction amounts to the case where $\sigma_{xx} = \sigma$ and all other $\sigma_{ij} = 0$, or equivalently that

$ \scriptscriptstyle{\begin{pmatrix} \epsilon_{xx} \\ \epsilon_{yy} \\ \epsilon_{zz} \\ \epsilon_{yz} \\ \epsilon_{zx} \\ \epsilon_{xy} \end{pmatrix} = \begin{pmatrix} \sigma (\frac{C_{11}C_{33}-C_{13}^2}{\alpha(C_{11}-C_{12})})  \\ \sigma (\frac{C_{13}^2-C_{12}C_{33}}{\alpha(C_{11}-C_{12})} ) \\ -\sigma (\frac{C_{13}}{\alpha}) \\ 0 \\ 0 \\ 0 \ \end{pmatrix}}$.

For the more complicated case of a uniaxial stress applied in the [110] direction, $\sigma_{xx} = \sigma_{yy} = \sigma_{xy} =  \frac{\sigma}{2}$ and all other $\sigma_{ij} = 0$.  This result is taken from \cite{Sun10}, but it can be seen from two different vantage points: 1) by resolving the applied stress into its $x$- and $y$-components while accounting for the enlarged cross sectional area of the plane normal to the [110] direction, or 2) by applying a 45$^{\circ}$ orthogonal coordinate transformation to the stress tensor.  Using this result, we obtain that

$ \scriptscriptstyle{\begin{pmatrix} \epsilon_{xx} \\ \epsilon_{yy} \\ \epsilon_{zz} \\ \epsilon_{yz} \\ \epsilon_{zx} \\ \epsilon_{xy} \end{pmatrix} = \begin{pmatrix}  \frac{\sigma}{2} ( \frac{C_{11}C_{33}-C_{13}^2}{\alpha(C_{11}-C_{12})} + \frac{C_{13}^2-C_{12}C_{33}}{\alpha(C_{11}-C_{12})} ) \\ \frac{\sigma}{2} ( \frac{C_{11}C_{33}-C_{13}^2}{\alpha(C_{11}-C_{12})} + \frac{C_{13}^2-C_{12}C_{33}}{\alpha(C_{11}-C_{12})} ) \\ -\sigma ( \frac{C_{13}}{\alpha} )  \\ 0 \\ 0 \\ \frac{\sigma}{2} ( \frac{1}{C_{66}} ) \end{pmatrix} = \begin{pmatrix}  \frac{\sigma}{2} ( \frac{C_{33}}{\alpha} ) \\ \frac{\sigma}{2} ( \frac{C_{33}}{\alpha} ) \\ -\sigma ( \frac{C_{13}}{\alpha} ) \\ 0 \\ 0 \\ \frac{\sigma}{2} ( \frac{1}{C_{66} )} \end{pmatrix}}$

\noindent
At this stage, we can now plug in numbers to estimate the strain required to move across the HO-LMA phase boundary.  For $(C_{11}, C_{12}, C_{13}, C_{33}, C_{44}, C_{66}) = (255, 48, 86, 313, 133, 188)$ GPa (taken from \cite{wolf_1994}, which are extrapolated to $T = 0$ K), a stress of magnitude $\sigma = 0.6$ GPa applied in the [100] direction induces a strain of

$\scriptscriptstyle{\begin{pmatrix} \epsilon_{xx} \\ \epsilon_{yy} \\ \epsilon_{zz} \\ \epsilon_{yz} \\ \epsilon_{zx} \\ \epsilon_{xy} \end{pmatrix} = \begin{pmatrix} 2.62 \times 10^{-3} \\ -2.76 \times 10^{-4} \\ -6.45 \times 10^{-4} \\ 0 \\ 0 \\ 0 \end{pmatrix}}$

\noindent
while a stress of magnitude $\sigma = 0.6$ GPa applied in the [110] direction induces a strain of

$\scriptscriptstyle{\begin{pmatrix} \epsilon_{xx} \\ \epsilon_{yy} \\ \epsilon_{zz} \\ \epsilon_{yz} \\ \epsilon_{zx} \\ \epsilon_{xy} \end{pmatrix} = \begin{pmatrix} 1.17 \times 10^{-3} \\ 1.17 \times 10^{-3} \\ -6.45 \times 10^{-4} \\ 0 \\ 0 \\ 1.60 \times 10^{-3}  \end{pmatrix}}$

\noindent
This analysis reveals that the strain in the [100] ([110]) direction associated with a uniaxial stress of 0.6 GPa in the [100] ([110]) direction is of order $10^{-3}$. In contrast, the largest strain experienced by the samples in our experiments corresponding to a full voltage sweep of the piezo stack, is of order $\frac{\Delta l}{l} \sim 10^{-4}$. Some of the transverse strains associated with the uniaxial stress are nevertheless smaller than $10^{-3}$, so the analysis described in the main text to extract the elastoresistivity coefficients was repeated for a smaller range of strains. The temperature dependence of the elastoresistivity coefficients was the same for both small ($\sim 10^{-5}$) and large ($\sim 10^{-4}$) strain regimes, consistent with the linear response observed for the resistance as a function of strain (see for example Figure 1 of the main text). The only difference in the temperature dependence of the elastoresistivity coefficients for the two strain ranges was that the signal-to-noise became worse for the smaller strain range. These effects are discussed in more detail in Section \ref{sec:D}.

\begin{figure}
\includegraphics[width=8.5cm]{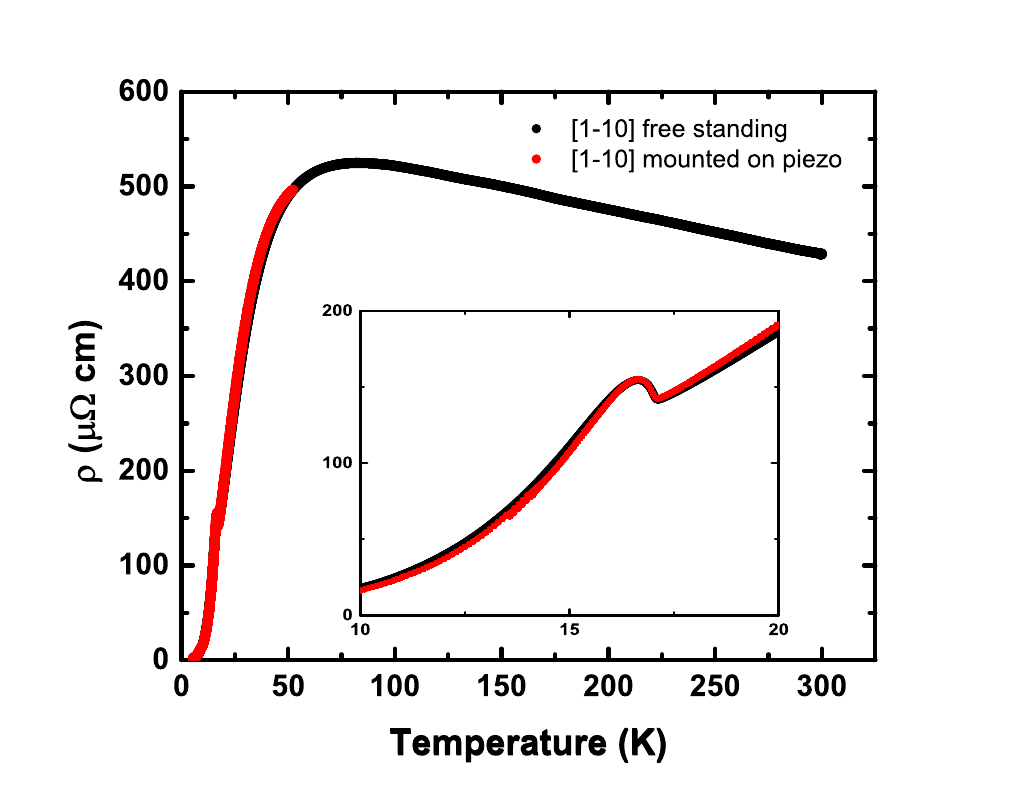} 
\caption{(Color online) Resistivity as a function of temperature on [1$\bar{1}$0] Sample 3, both free-standing and glued to the piezo stack. The two transport curves lay on top of one another, indicating the effect of built-in strain is negligible, i.e. $R(V=0) = R_0$.}
\label{figure_freevpiezo}
\end{figure}

Another source of strain comes from mounting the sample to the piezo stack, which creates a ``built-in'' strain from the epoxy \cite{Butkovicova13}. To test for the effects of built-in strain, the resistivity as a function of temperature was measured with the sample free-standing and when glued to the piezo stack (Figure \ref{figure_freevpiezo}). The two transport curves are essentially identical, showing no discernible shift in the Hidden Order transition temperature. Therefore, we conclude that any built-in strain from the epoxy does not contribute to the resistive response, i.e., $\frac{R(V)-R_0}{R_0} = \frac{R(V)-R(V=0)}{R(V=0)}$, where $R_0$ is the free standing resistance and R(V=0) is the resistance of the sample mounted on the piezo at zero applied voltage.

\subsection{Curve Fitting}

Neglecting the sharp anomaly at $T_{HO}$, the elastoresistivity coefficient $2m_{66}$ exhibits a monotonic increase with decreasing temperature from 60 K (the highest temperature at which there is a measureable elastoresistance) down to just above $T_{HO}$. In the following two sections, we briefly describe the procedures used to fit the data first to Curie-Weiss temperature-dependence (motivated by the Ginzburg-Landau models described in Section I), and then to a more general power law.

\subsubsection{Curve Fitting - Curie-Weiss}

\begin{figure}
\includegraphics[width=8.5cm]{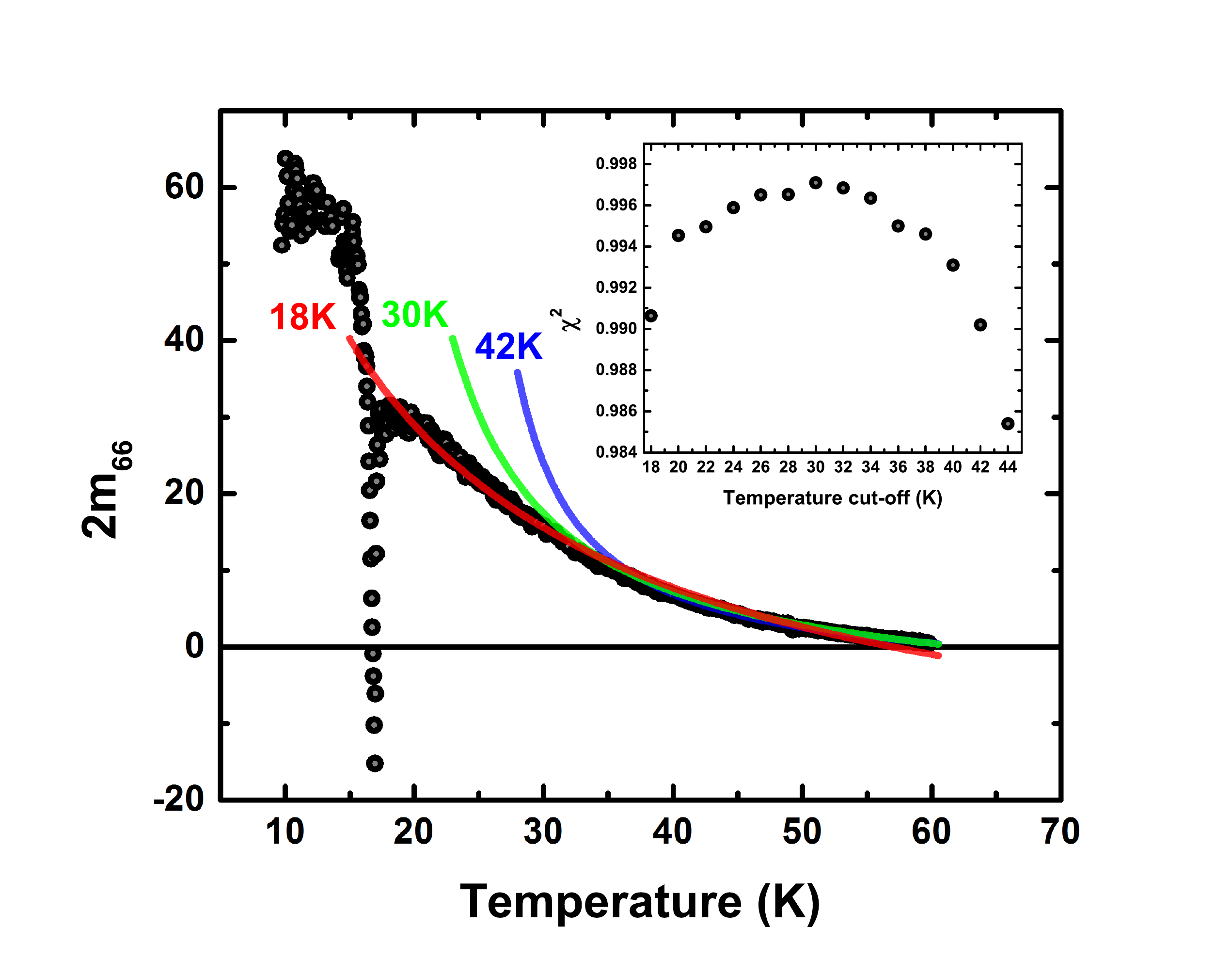} 
\caption{(Color online) A plot of the elastoresistivity coefficient $2m_{66}$ as a function of temperature. The data were fit to \eqref{eq:36} with $\alpha=1$. The red curve is a fit from 18 K to 60 K, the green from 30 K to 60 K, and the blue from 42 K to 60 K. The inset shows the reduced $\chi^2$ for each fitting regime. The best fit is over the region 30 K to 60 K, as defined by $\chi^2_{\textrm{red}}$ closest to 1.}
\label{figure_cutoff}
\end{figure}

Motivated by the Ginzburg-Landau theory described in Section I and in the main text, the $2m_{66}$ elastoresistance data were fit to a Curie-Weiss temperature dependence $2m_{66} = \frac{C}{T-\theta} + 2m_{66}^{0}$. When the data are fit over the full temperature range from  60 K down to  18 K (i.e. to just above the anomaly centered at 17.2 K), a clear systematic deviation can be seen by inspection (red curve of Figure \ref{figure_cutoff}), which is confirmed by the reduced $\chi^2$ value (inset Figure \ref{figure_cutoff}). Similar fits were performed between 60 K and different lower-bound cut-off temperatures to find a reduced $\chi^2$ which was closest to 1, which is how we define the best fit (inset to Figure \ref{figure_cutoff}). The best fit determined by this method is found to be for the range between 60 K and 30 K (green curve, Figure \ref{figure_cutoff}), with fit parameters given in the main text.

\subsubsection{Curve Fitting - Power Law}

In addition to fitting the data to a Curie-Weiss temperature dependence, as described above, a more general fit to an arbitrary power law was also attempted. However, with less than a decade in reduced temperature, the data are not well-suited to such a scaling analysis, and the results  are not conclusive. We include this analysis here for completeness. 

We start by assuming that the elastoresistance is given by a single power law:

\begin{eqnarray}
\label{eq:36}
2m_{66} = 2m_{66}^{0} + \frac{C}{(T-\theta)^{\alpha}}
\end{eqnarray}

\noindent where $2m_{66}^0$, $C$, $\theta$ and $\alpha$ are treated as fit parameters. Importantly, $\alpha$ is a nonlinear fit parameter, which greatly complicates the problem. One can linearize $\alpha$ (the parameter we are most interested in) by subtracting off $2m_{66}^{0}$, taking the natural log of both sides, and then taking a temperature derivative.  If we could precisely estimate $2m_{66}^{0}$ and sensibly smooth the data in taking the temperature derivative, then we would be left with the simple task of fitting

\begin{eqnarray}
\label{eq:37}
\frac{d\ln(2m_{66} - 2m_{66}^{0})}{dT} = \frac{-\alpha}{T-\theta}
\end{eqnarray}

The problem with the procedure leading to (\ref{eq:37}) is that if we misestimate $2m_{66}^{0}$ by some
amount $\delta$, then we no longer have a linearized fit form for $\alpha$:

\begin{eqnarray}
\label{eq:38}
\ln(2m_{66} - 2m_{66}^{0}) = \ln \left( \delta + \frac{C}{(T-\theta)^{\alpha}} \right).
\end{eqnarray}

\noindent Furthermore, there is no sensible way to linearize (\ref{eq:38}) via a Taylor expansion since we expect the temperature dependent term to dominate near criticality but to progressively approach zero at higher temperatures. These issues and others have been addressed in the literature \cite{Sobotta85}.

Instead, we chose a graphical method.  Assuming the functional form (\ref{eq:36}) and fixing a particular $\alpha$, one can unbiasedly estimate the other three linear parameters (and in particular the temperature independent term $2m_{66}^{0}$) through a normal fitting routine. Since (\ref{eq:36}) is equivalent to

\begin{eqnarray}
\label{eq:39}
\frac{1}{(2m_{66}-2m_{66}^{0})^{1 / \alpha}} = \frac{T-\theta}{C^{1 / \alpha}}
\end{eqnarray}

\noindent it is straightforward to compute $\frac{1}{(2m_{66}-2m_{66}^{0})^{1 / \alpha}}$ and to
graph the result versus temperature.  Doing this for many values of $\alpha$ yields the critical
exponent, as the right choice will result in a quantity linearly proportional to temperature.

Importantly, this method avoids the pitfalls of nonlinear parameter estimation and numerical
differentiation; however, it is still quite susceptible to errors in $2m_{66}^{0}$, as any error
$\delta$ results in the highly nonlinear expression

\begin{eqnarray}
\label{eq:40}
\frac{1}{(2m_{66}-2m_{66}^{0})^{1 / \alpha}} = \frac{1}{(\delta + \frac{C}{(T-\theta)^{\alpha}})^{1 / \alpha}}
\end{eqnarray}

\begin{figure}
\includegraphics[width=8.5cm]{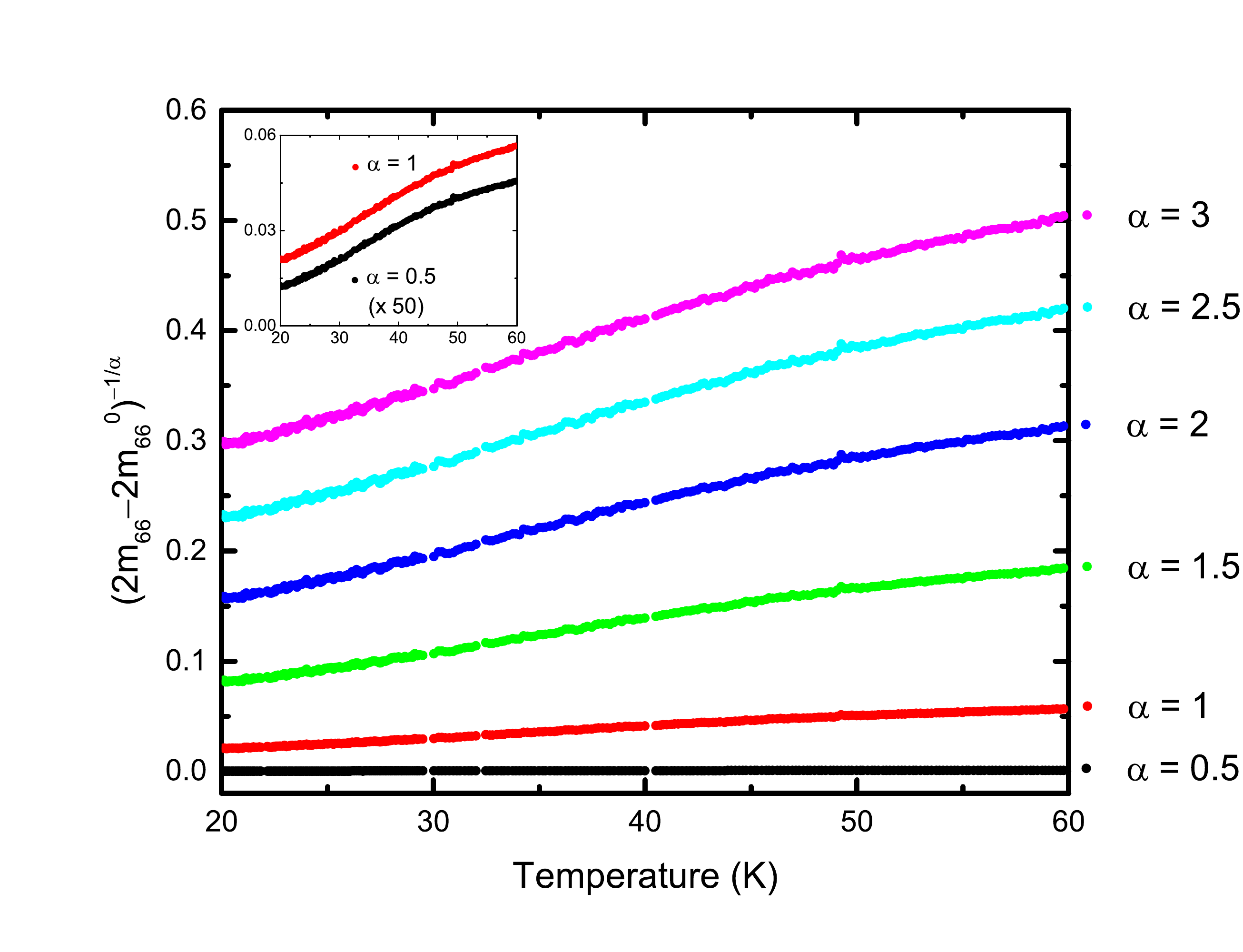}
\caption{\label{fig:powerlawfitting} Inverse divergent elastoresistivity coefficient versus temperature.
For the case where $2m_{66}$ obeys a power law divergence with a critical exponent $\alpha$, we should
observe a curve that is linear in temperature.  This linear relationship is not observed for any $\alpha$
spanning the range $0.5 \leq \alpha \leq 3$.  Inset: Zoom-in version on the $\alpha = 0.5$ and $\alpha = 1$
curves, highlighting nonlinear behavior.  The $\alpha = 0.5$ curve is rescaled by a factor of 50 for
illustrative purposes.}
\label{powerlawfitting}
\end{figure}

\noindent The nonlinearity from misestimating $2m_{66}^{0}$ expressed in (\ref{eq:40}) is worth keeping in mind
since $\frac{1}{(2m_{66}-2m_{66}^{0})^{1 / \alpha}}$ \emph{is not} linear in temperature for any $\alpha$ spanning the range $0.5 \leq \alpha \leq 3$ (Figure \ref{fig:powerlawfitting}). The fits are done over the temperature range $20$ K $\leq T \leq 60$ K, where we are limited on the low end by fluctuations above $T_{HO}$ and on the high end by our ability to resolve the elastoresistive response.  Note that the estimate of $2m_{66}^{0}$ is itself dependent on the particular $\alpha$ chosen, and so varies between curves.  Higher values of $\alpha$ yield highly unphysical estimates of $\theta$ and are hence neglected.

From the temperature dependence of the $2m_{66}$ data (Figure \ref{powerlawfitting}), we do not unequivocally observe a power law divergence with an extractable critical exponent. The task of fitting critical exponents is potentially complicated by our limited range in temperature (less than a decade), and this might be why it is difficult to distinguish between different $\alpha$'s. The results are also potentially skewed by incorrect estimations in the temperature
independent parameter $2m_{66}^{0}$. Consequently, the most physically meaningful fit to the data is provided by the Curie-Weiss model described in the main text (i.e. fixing the exponent $\alpha=1$).

\subsection{Nonlinear Elastoresistance near $T_{HO}$}\label{sec:D}

\begin{figure}
\includegraphics[width=6.5cm]{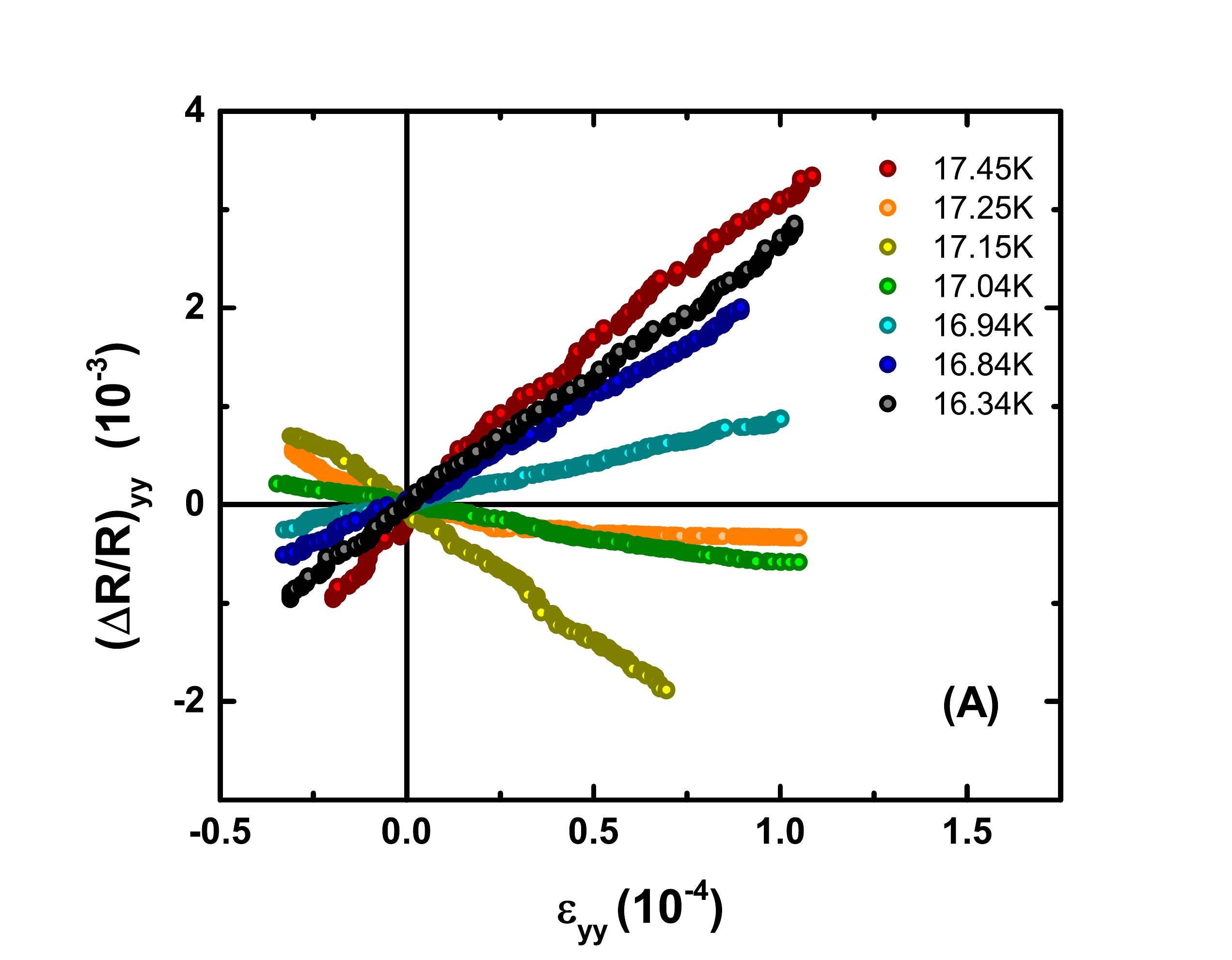} 
\includegraphics[width=6.5cm]{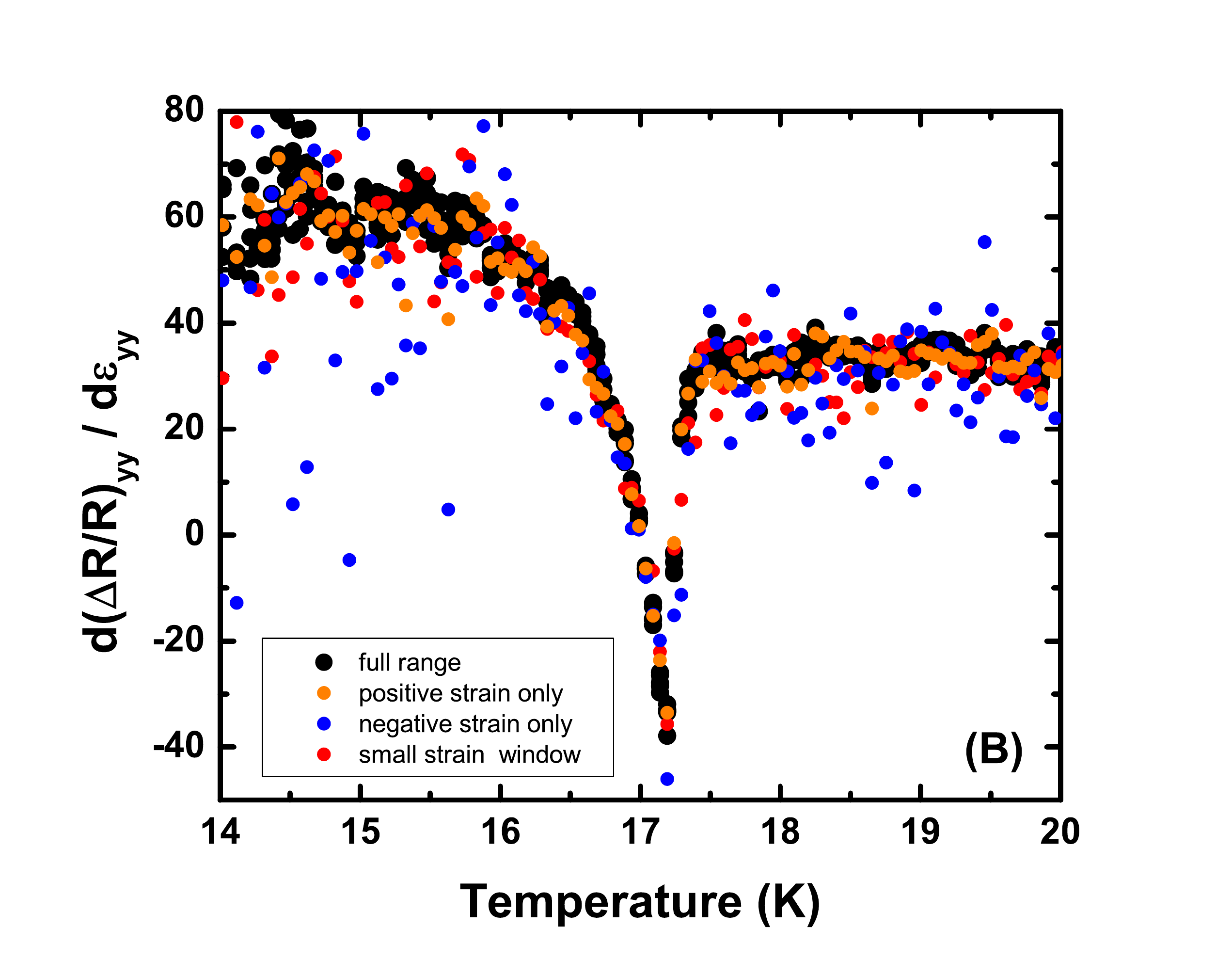}
\includegraphics[width=6.5cm]{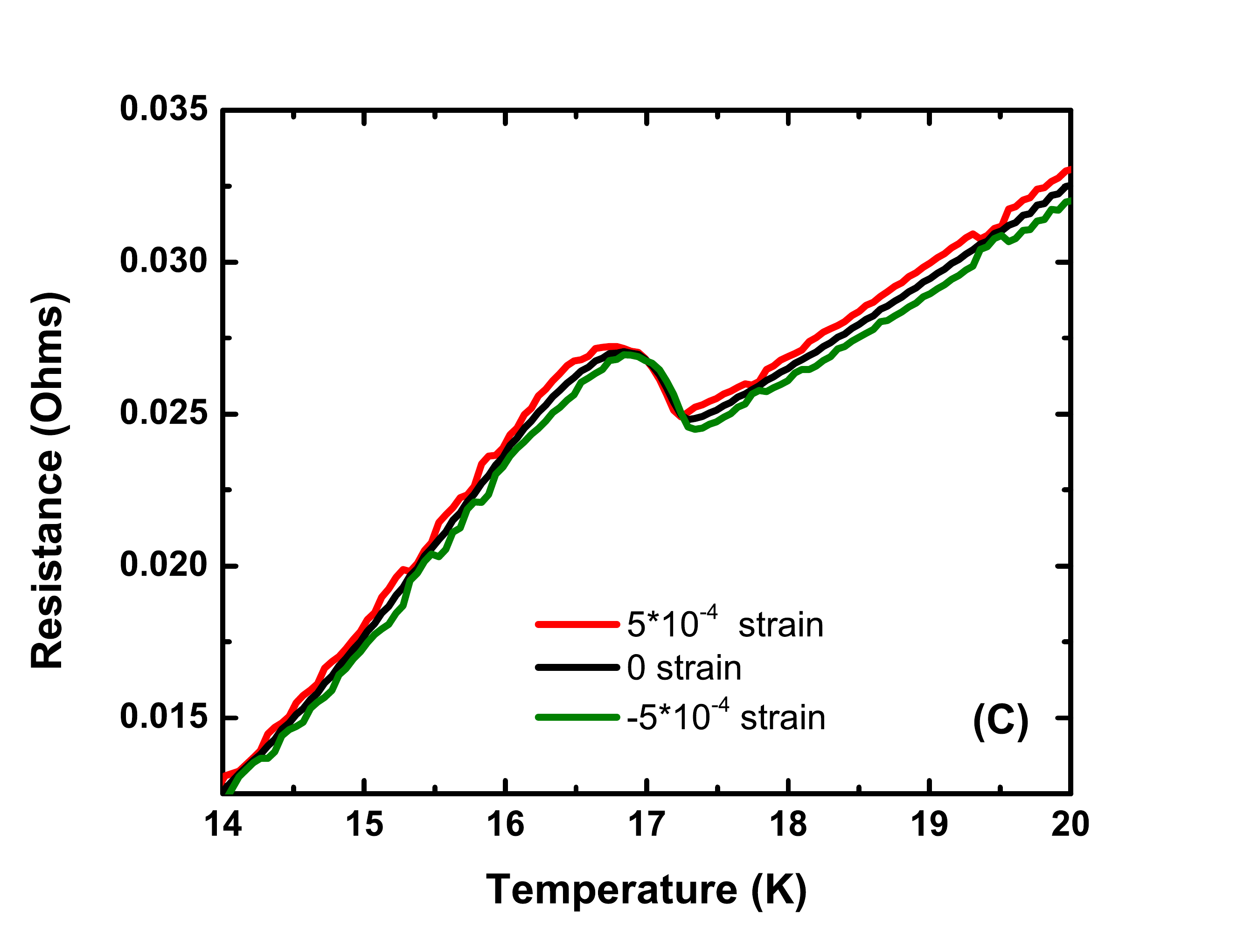}
\caption{(Color online) (a) $(\frac{\Delta R}{R})_{yy}$ as a function of applied strain $\epsilon_{yy}$ around the Hidden Order transition for Sample 3 ([1$\bar{1}$0]). There is a clear sign change in the slope of $(\frac{\Delta R}{R})_{yy}$ at $T_{HO}$ = $17.15$ K. In a small temperature range ($\pm 0.15$ K) about $T_{HO}$, the resistive response to applied strain is nonlinear. The slopes in (a) are fit and plotted in (b) for different strain regions. Black points refer to the full strain region of $-50$ V to $+150$ V (which correspond to strains $\sim$ $-0.25 \times 10^{-4}$ to $1 \times 10^{-4}$). Orange points in (b) are from slopes fit over a $0$ V to $150$ V strain region in (a) (strains of $\sim$ $0$ to $1 \times 10^{-4}$). Blue points in (b) are from slopes fit over a $-50$ V to $0$ V strain region in (a) (strains of $\sim$ $-0.25 \times 10^{-4}$ to $0$). Red points in (b) are from slopes fit over a $-50$ V to $50$ V region in (a) (strains of $\sim$ $-0.25 \times 10^{-4}$ to $0.25 \times 10^{-4}$). Comparing the black points and the red points in (b), it can be seen that the magnitude of the elastoresistivity coefficients are slightly underestimated if the data are fit to a linear function over the entire strain range and overestimated if fitting only the negative strain region. In (c) we plot resistance as a function of temperature for a fixed strain in the [1$\bar{1}$0] sample (see main text for further details).  From the shift in the point of inflection in the resistance, application of $+ 5 \times 10^{-4}$ strain (red curve) and of $- 5 \times 10^{-4}$ strain (green curve) shifts the transition temperature $T_{HO}$ by approximately 100 mK per $10^{-3}$ of strain.}
\label{figure_strain_THO}
\end{figure}

For temperatures more than $\sim$150 mK away from $T_{HO}$, the elastoresistance is linear with strain over the entire range of strain experienced by the material in our experiments (which, in the vicinity of $T_{HO}$ is approximately $-0.5 \times 10^{-4} < \epsilon_{yy} < +1 \times 10^{-4}$); however, very close to $T_{HO}$ = 17.15 K, the elastoresistance develops a nonlinear response to strain. This effect can be seen in Figure \ref{figure_strain_THO}(a), which shows successive measurements of $(\frac{\Delta R}{R})_{yy}$ for a [1$\bar{1}$0] oriented crystal as a function of $\epsilon_{yy}$ for temperatures spanning $T_{HO}$. In the nonlinear regime, the magnitude of the elastoresistivity coefficents are slightly \emph{underestimated} if the data are fit to a linear function over the entire data range. This effect can be seen in Figure \ref{figure_strain_THO}(b), which shows $(\frac{\Delta R}{R})_{yy}$/$\epsilon_{yy}$ evaluated for different strain ranges (see Figure \ref{figure_strain_THO} caption for further detail). It is important to note, however, that the sign change of the slope of the elastoresistance is a robust feature, observed for large and small strains and for temperatures that extend beyond the nonlinear regime.  

The physical origin of the nonlinear response very close to $T_{HO}$ is unclear. The effect could be associated with critical behavior, or possibly with strain-induced changes in $T_{HO}$. To investigate the latter idea, we estimate the rate at which $T_{HO}$ is affected by anisotropic strain. We do this by extracting the temperature dependence of the resistivity for fixed values of the strain. This is obtained by performing a linear fit to the elastoresistance over different strain ranges (see Figure \ref{figure_strain_THO} caption for further detail) and using the fit parameter to calculate the resistance for a specific strain (we choose $\pm 5 \times 10^{-4}$, but the exact value is immaterial). The resulting curves are shown in Figure \ref{figure_strain_THO}(c), together with the temperature dependence of the resistivity for the same unstrained crystal (mounted on the piezo, with 0 V applied to the piezoelectric stack). For all temperatures more than $\sim$150 mK from $T_{HO}$, the resistance of the strained sample is independent of the fit range used to calculate the elastoresistivity coefficients, consistent with the linear elastoresistive response. Inspection of the curves in Figure \ref{figure_strain_THO}(c) indicates that $T_{HO}$ increases (decreases) for negative (positive) strains $\epsilon_{yy}-\epsilon_{xx}$. This effect is consistent with expectations for a phase transition involving a multi-component order parameter and with the model developed in Section I, as can be appreciated by inspection of Equation \eqref{eq:7}.  The analysis indicates a change in $T_{HO}$ of approximately $100$ mK per $10^{-3}$ strain. Close to $T_{HO}$, such a strain-induced change in $T_{HO}$ would yield a nonlinear response, possibly accounting for some of the nonlinearity at $T_{HO}$. Significantly, for smaller strains, the change in $T_{HO}$ would be correspondingly smaller. Since the anomaly in $2m_{66}$ is observed for large and small strain ranges alike (Figure \ref{figure_strain_THO}(b)), and since the response is linear for all temperatures more than $\sim$150 mK from $T_{HO}$ (in comparison to the width of the anomaly, which extends for over 1000 mK), we conclude that the anomaly itself is not primarily associated with strain-induced changes in $T_{HO}$. Nevertheless, the nonlinear effects described above preclude a critical scaling analysis, at least for the strains employed in the current study.

\begin{references}

\bibitem{palstra_1985} T. T. M. Palstra, A. A. Menovsky, J. van den Berg, A. J. Dirkmaat, P. H. Kes, G. J. Nieuwenhuys, and J. A. Mydosh, Phys. Rev. Lett. \textbf{55}, 2727 (1985).

\bibitem{schlabitz_1986} W. Schlabitz, J. Baumann, B. Politt, U. Rauchschwalbe, H.M. Mayer, U. Ahlheim, and C.D. Bredl, Z. Phys. \textbf{62}, 171 (1986).

\bibitem{maple_1986} M. B. Maple, J. W. Chen, Y. Dalichaouch, T. Kohara, C. Rossel, M. S. Torikachvili, M. W. McElfresh, and J. D. Thompson, Phys. Rev. Lett. \textbf{56}, 185 (1986).

\bibitem{JM_2011} J. A. Mydosh and P. M. Oppeneer, Rev. Mod. Phys. \textbf{83}, 1301 (2011).

\bibitem{kristen_2009} Kristjan Haule and Gabriel Kotliar, Nature Physics \textbf{5}, 796 (2009).

\bibitem{varma_2006} C. M. Varma and Lijun Zhu, Phys. Rev. Lett. \textbf{96}, 036405 (2006).

\bibitem{flint_2013} Premala Chandra, Piers Coleman, and Rebecca Flint, Nature \textbf{493}, 621 (2013).

\bibitem{das_2014} Tanmoy Das, Phys. Rev. B \textbf{89}, 045135 (2014).

\bibitem{matsuda_2012} Hiroaki Ikeda, Michi-To Suzuki, Ryotaro Arita,	 Tetsuya Takimoto, Takasada Shibauchi, and Yuji Matsuda, Nature Physics \textbf{8}, 528 (2012).

\bibitem{kotliar_2010} K. Haule and G. Kotliar, EPL \textbf{89}, 57006 (2010). 

\bibitem{tripathi_2002} P. Chandra, P. Coleman, J. A. Mydosh, and V. Tripathi, Nature \textbf{417}, 831 (2002).

\bibitem{mydosh_2009} S. Elgazzar, J. Rusz, M. Amft, P. M. Oppeneer, and J. A. Mydosh, Nature Materials \textbf{8}, 337 (2009).

\bibitem{torque_2011} R. Okazaki, T. Shibauchi, H. J. Shi, Y. Haga, T. D. Matsuda, E. Yamamoto, Y. Onuki, H. Ikeda, and Y. Matsuda, Science \textbf{331}, 439 (2011).

\bibitem{footnote_1} Recent high resolution X-ray measurements reveal a small but apparently abrupt orthorhombicity onsetting at $T_{HO}$ \cite{taka_private}, which suggests that the phase transition may be first order. The large mean field-like heat capacity anomaly, however, validates the assumption of a continuous (or, at worst, weakly first order) transition, which is the view that we adopt here.  In this case, the measurements reported here can reasonably be interpreted within a framework of a fluctuational regime present above $T_{HO}$.

\bibitem{taka_private} Private communication, Takasada Shibauchi.

\bibitem{footnote_2} i.e., corresponding to cases in which the tetragonal lattice is stretched along either the [100] or [010] direction, or along the [110] or [1$\bar{1}$0] direction respectively, corresponding to $B_{1g}$ and $B_{2g}$ representations of the $D_{4h}$ point group.

\bibitem{footnote_3} As defined, this is the \emph{bare} nematic susceptibility, unrenormalized by interaction with the crystal lattice. [See ``Settling the Chicken and Egg question in iron based superconductors'', Journal Club for Condensed Matter Physics; http://www.condmatjournalclub.org/?p=2247].

\bibitem{JH_2012} J.-H. Chu, H.-H. Kuo, J. G. Analytis, and I. R. Fisher, Science \textbf{337}, 710 (2012).

\bibitem{HH_2011} H.-H. Kuo, J.-H. Chu, Scott C. Riggs, L. Yu, P. L. McMahon, K. De Greve, Y. Yamamoto, J. G. Analytis, and I. R. Fisher, Phys. Rev. B \textbf{84}, 054540 (2011).

\bibitem{HH_2013} H.-H. Kuo, Maxwell C. Shapiro, Scott C. Riggs, I. R. Fisher, Phys. Rev. B \textbf{88}, 085113 (2013).

\bibitem{SOM} See Supplemental Material.

\bibitem{footnote_4} In the example cited of the unidirectional density wave, the nematic order parameter $\mathcal{N}$ is parasitic to the density wave order. However, the same terms in the Ginzburg-Landau analysis would occur if the nematic order parameter existed independently of the density wave order. So long as the bare mean field critical temperature associated with the nematic order in the associated theory is below that of the density wave order, there is only ever one phase transition at which the unidirectional density wave order onsets, and the temperature dependence of the nematic susceptibility is the same in both cases, regardless of the physical origin of the nematic order parameter (see supporting material for more details \cite{SOM}).

\bibitem{footnote_5} The notion of a ``multicomponent'' order parameter depends in turn on whether the Hidden Order phase preserves or breaks lattice translation symmetry (corresponding to a $\vec Q=0$ and finite $\vec Q$ order parameters, respectively).  In the former case, a multicomponent order parameter must belong to a multidimensional irreducible representation (irrep) of the space group---examples are the $E_g$ and $E_u$ irreps of the $D_{4h}$ point group.  There are two possible realizations of multicomponent order parameters at finite $\vec Q$: (1) the order parameter is a scalar but can order in one of two directions  $Q_x$ and $Q_y$ (e.g., unidirectional density wave order).  The corresponding multicomponent order parameter is the set ${\Delta(Q_{x}), \Delta(Q_{y})}$. The specific example cited has $B_{1g}$ symmetry, and there is an equivalent case with $B_{2g}$ symmetry (unidirectional density wave running along [110] and [1$\bar{1}$0] directions); (2) the finite $\vec Q$ order itself belongs to a multidimensional irrep such as the $E_g$ group. For this case, $\vec Q$ could be a high symmetry point ($D_{4h}$ examples are $Q = (0,0, \ell)$ or $(\pi, \pi, \ell)$, where $\ell$ represents a modulation along the z-axis).

\bibitem{deltaR} ($\frac{\Delta \rho}{\rho})_i = \sum\limits_{j=1}^6 m_{ij}\epsilon_{j}$, where $1 = xx, 2 = yy, 3 = zz, 4 = yz, 5 = zx, 6 = xy$

\bibitem{NMR_2013} K. R. Shirer, J. T. Haraldsen, A. P. Dioguardi, J. Crocker, N. apRoberts-Warren, A. C. Shockley, C.-H. Lin, D. M. Nisson, J. C. Cooley, M. Janoschek, K. Huang, N. Kanchanavatee, M. B. Maple, M. J. Graf, A. V. Balatsky, and N. J. Curro, Phys. Rev. B \textbf{88}, 094436 (2013).

\bibitem{ES_2014} Private communication, Elizabeth Schemm and Aharon Kapitulnik.


\bibitem{MF_1968} Michael E. Fisher and J. S. Langer, Phys. Rev. Lett. \textbf{20}, 665 (1968). 

\bibitem{Cp_eric} The phonon component to the specific heat was subtracted out by using ThRu$_{2}$Si$_{2}$, such that $C_{elec}$(URu$_2$Si$_2$) = [$C$(URu$_2$Si$_2$) $-$ $C$(ThRu$_2$Si$_2$)].


\bibitem{wolf_1994} B. Wolf, W. Sixl, R. Graf, D. Finsterbusch, G. Bruls, B. L\"{u}thi, E. A. Knetsch, A. A. Menovsky, and J. A. Mydosh, J. Low Temp. Phys. \textbf{94}, 307 (1994). 


\bibitem{yanagisawa_2011} T. Yanagisawa, H. Saito, Y. Watanabe, Y. Shimizu, H. Hidaka, and H. Amitsuka, J. Phys. Conf. Ser. \textbf{391}, 012079 (2012).

\bibitem{footnote_6} In addition to the weak coupling to the crystal lattice, we note that the weakly first order nature of the phase transition suggested by recent X-ray diffraction experiments \cite{taka_private} will cut off the critical fluctuations for small values of the reduced temperature, possibly accounting for the absence of any appreciable renormalization of $C_{66}$ through the Hidden Order phase transition.

\bibitem{Fernandes12} R. M. Fernandes, A. V. Chubukov, J. Knolle, I. Eremin, and J. Schmalian, Phys. Rev. B \textbf{85}, 024534 (2012). 

\bibitem{Ballato95} Arthur Ballato, ``Poisson's Ratio for Tetragonal Crystals'', Army Research Laboratory, 1995.

\bibitem{Sun10} Y. Sun, S. E. Thompson, and T. Nishida, \textit{Strain Effect in Semiconductors: Theory and Device Applications}, ``Chapter 2: Stress, Strain, Piezoresistivity, and Piezoelectricity'', Springer, 2010.

\bibitem{Butkovicova13}  D. Butkovicova, X. Marti, V. Saidl, E. Schmoranzerova-Rozkotova, P. Wadley, V. Holy, and P. Nemec, Rev. Sci. Instr. \textbf{84}, 103902 (2013).

\bibitem{Sobotta85} G. Sobotta, J. Phys. C: Solid State Phys. \textbf{18}, 2065 (1985).


\end{references}
\end{document}